\documentclass[12pt,peerreview,letterpaper]{IEEEtran}
\usepackage{graphicx}
\usepackage{amsmath}
\usepackage{enumerate}
\usepackage{graphicx}
\usepackage{setspace}
\usepackage{filecontents}
\usepackage{amsmath,amssymb,multirow,epsfig} 
\usepackage{epsfig}
\usepackage{array}
\usepackage{psfrag}
\usepackage{pstricks,pst-node,pst-text,pst-3d,pst-plot}
\usepackage{trig}
\usepackage{amssymb}
\usepackage{anysize}
\usepackage{color}
\usepackage{cite}
\usepackage{tikz}
\usepackage{amsmath}

\usepackage{graphicx}
\include{psfig}
\usepackage{epsfig}
\usepackage{array}
\usepackage{psfrag}
\usepackage{pstricks,pst-node,pst-text,pst-3d,pst-plot}
\usepackage{trig}
\usepackage{amssymb}
\usepackage{anysize}
\usepackage{color}
\usepackage{cite}
\usepackage{tikz}
\usepackage{amsmath}
\usepackage{float}
\usepackage{amsthm}
\theoremstyle{remark}

\theoremstyle{lemma}
\newtheorem{lemma}{Lemma}[section]

\newcommand{\bx}{{\bf{x}}}

\newcommand{\bw}{{\bf{w}}}
\newcommand{\bW}{{\bf{W}}}
\newcommand{\bof}{{\bf{f}}}
\newcommand{\bF}{{\bf{F}}}
\newcommand{\bh}{{\bf{h}}}

\newcommand{\ba}{{\bf{a}}}
\newcommand{\bA}{{\bf{A}}}
\newcommand{\bb}{{\bf{b}}}
\newcommand{\bB}{{\bf{B}}}

\newcommand{\bP}{{\bf{P}}}
\newcommand{\bQ}{{\bf{Q}}}
\newcommand{\bc}{{\bf{c}}}
\newcommand{\br}{{\bf{r}}}
\newcommand{\bt}{{\bf{t}}}
\newcommand{\bT}{{\bf{T}}}
\newcommand{\bI}{{\bf{I}}}
\newcommand{\bk}{{\bf{k}}}

\newcommand{\bOmega}{{\boldsymbol{\Omega}}}

\newcommand{\bDelta}{{\boldsymbol{\Delta}}}
\newcommand{\diag}{{\rm{diag}}}

\renewcommand{\baselinestretch}{1.5}
\begin{document}

\title{Optimal Relaying Beamforming in Multiple Access Broadcast Channel (MABC) Bidirectional Cognitive Radio Networks in Presence of Interferers}



%
%
%

\author{
   \IEEEauthorblockN{
   Mohammad Zaeri-Amirani\IEEEauthorrefmark{1},
    Fatemeh Afghah\IEEEauthorrefmark{1},
    and
     Jonathan D.~ Ashdown\IEEEauthorrefmark{2}}
   \IEEEauthorblockA{
     \IEEEauthorrefmark{1}School of Informatics, Computing and Cyber Systems,
    Northern Arizona University, Flagstaff, AZ, United States\\
    Email: \{Mohammad.Zaeri-Amirani,fatemeh.afghah\}@nau.edu \\
    }
   \IEEEauthorblockA{
    \IEEEauthorrefmark{2}Air Force Research Laboratory,
    Rome, NY, United States\\
     Email: jonathan.ashdown@us.af.mil
   }
 }




\maketitle

\begin{abstract}
In this paper, a general cognitive radio system consisting of a set of users with different level of spectrum access including two primary transceivers and several types of secondary users is considered. It is assumed that two secondary users operate based on an underlay model at the same frequency bandwidth and at the same time as the primary users based on a multiple access broadcast channel (MABC) bidirectional beamforming scheme. Other secondary users provide a relaying service to the primary users in exchange for the opportunity to send their messages towards their own destinations for a fixed portion of the communication cycle. In addition, it is assumed that some interferers are active during the communication cycle and cause interference for the network. Furthermore, it is assumed that only partial channel state information (CSI) between interferers and other nodes in the network is available.
 We provide a robust optimization method against imperfection on the interferers' CSI to maximize the joint primary and secondary {signal-to-interference-plus-noise-ratio (SINR)} with the assumption of limited available power at the secondary relays. An amplify-and-forward (AF) relaying scheme is deployed at the secondary relays and the optimal beamforming is obtained using second order convex programming (SOCP) method. The simulation results show the performance of the proposed beamforming method against the existence of interferers, and demonstrate the effectiveness of our robust method against uncertainty in knowledge of interferers' CSIs \footnote{This work is partially supported by Air Force Research Laboratory under grant number 18.0257. Distribution Statement A: Approved for Public Release; distribution unlimited: 88ABW-2017-5903 on 21 Nov. 2017.}.
 %
%

\end{abstract}
\index{Amplify-and-forward relaying, bidirectional relay networks, cognitive radio, imperfect channel state information, and robust optimization }

\section{Introduction}\label{Sec:Intro}
While the ever-increasing demand for wireless service makes the radio spectrum one of the most valuable and scarce resources for wireless communication, recent studies have shown that the spectrum is not efficiently utilized at some locations for certain times of the day \cite{spectrum2002spectrum}. Dynamic spectrum management is a new paradigm to manage the radio spectrum in a dynamic manner by allowing cognitive nodes to utilize the unused bandwidth \cite{wyglinski2008cognitive}. Cognitive radio systems are usually composed of legacy spectrum owners, primary users (PUs), and cognitive devices seeking to access the PU's spectrum, called secondary users (SUs).

Generally, dynamic spectrum technologies are broadly categorized into the two categories of \emph{common model} and \emph{property-right model} \cite{jovicic2009cognitive}. Unlike the common model for spectrum sharing, where the primary users are oblivious to the presence of SUs, in the property-right model, the PUs can willingly lease a portion of their spectrum to the SUs in exchange for monetary benefits or physical compensations. This compensation could be in form of providing relaying service, energy harvesting or cooperative jamming for the PUs  \cite{CR-Afghah2013,afghah2014cooperative,Simeone2008,Afghah_INFOCOM,afghahIJHCR}. The property-right model for spectrum sharing in exchange for relaying service, also known as 'cooperative spectrum leasing' has received much attention in the last years, as it offers a win-win solution for both licensed and unlicensed users. The primary users can benefit from such spectrum leasing by enhancing their quality of service (QoS), in particular when experiencing a poor channel condition, while the secondary users can obtain the chance of affordable spectrum access. Furthermore, the property-right model can result in less energy consumption for the unlicensed users compared to the spectrum sensing models where they need to constantly sense the PUs' spectrum looking for spectrum holes.


In this work, we consider a general model of cognitive radio networks with co-existence of several types of SUs that operate in different modes. This network consists of two PU transceivers and two SU transceivers which desire to exchange their signals with the help of available secondary relays. It is assumed that two underlay SUs are allowed to share the radio spectrum with the PUs provided that their interference at the PUs' receivers remains below an acceptable threshold.
In addition to the two SUs operating in the underlay model, we consider the existence of multiple SUs that are interested in obtaining the spectrum access in exchange for providing relaying service for the PUs, based on the property-right model. Such model can limit the potential undesired interference that can be caused to the PUs, as well as the level of interference among the underlay SUs. Enabling a cooperative spectrum leasing to other SUs through property-right model can extend the number of SUs that can get the chance of spectrum access while benefiting the PUs through the cooperative relaying.
To further enhance the efficiency of radio spectrum utilization in this network, a two-way cooperative communication scenario is utilized in this network.
In general, the two methods of time division broadcast (TDBC) and multiple access broadcast (MABC) are utilized in two-way communication scenarios \cite{gavili2015optimal,Ubaidulla2014,ma2013robust, nguyen2017wireless,li2017resource,iranpanah2016distributed,Pareek,Zhang,Alsharoa}. In spite of TDBC scheme, where the transceivers send their signals in different time-slots, in MABC protocol both transceivers transmit their signals simultaneously. Since the proposed cooperative spectrum sharing mechanism is designed for cases when the quality of the direct link for the primary users is low, we consider an MABC scheme as it outperforms the TDBC scheme in such conditions \cite{zaeri2012}.

One of the main concerns regarding the implementation of spectrum sharing solutions is combating the interference caused by simultaneous transmission of SUs with the PUs. Spectrum sharing networks are also vulnerable to the presence of unfriendly interferers that, despite the compliance of SUs, are not designated to respect the QoS requirement for the PUs. Unlike the intentional interferers (jammers), who intend to disturb the PUs' communication, the unfriendly interferers degrade the PUs' performance due to simultaneous transmission without PUs' consent. A potential example of these interferers can be the sensing-based secondary users that may interfer with the PUs' communication due to false detection of the PUs' presence or synchronization imperfections. A cognitive radio network must account for such burdens, imposed by either intentional interferers or false detection errors in the sensing process. Therefore, we study a scenario for co-existence of two PUs, and two underlay SUs operating in a two-way relaying system with multiple secondary relays when multiple unfriendy interferers exist.
When such interferers exist, the information related to their channel state information is not usually available to the cognitive radio network. This is due to the fact that there is no cooperation between the unfriendly interferers and the networks' centralized controller which can only obtain an imperfect knowledge of the CSI. Withal, this imperfection can be due to time delays or frequency offset between the reciprocal channels as well as inaccurate channel estimation \cite{wang2011robust}. In order to study the impact of the imperfect CSI of the interferers, we assume that the CSI of all other channels is perfectly known. This assumption can be easily justified due to the existing collaboration between the PUs and SUs and the secondary relays, where CSI could be directly fed back from node to node \cite{jovicic2009cognitive}. While no collaboration between the interferers and the primary and secondary users is imagined, other mechanisms can be used to estimate the CSI between them. For instance, this CSI can be measured by a band manager and be provided using finite bandwidth channels \cite{suraweera2010capacity}. Eventually, this mechanism will cause inaccuracy in the estimated CSI which should be considered in the design of dynamic spectrum sharing systems.


To the best of our knowledge, this is the first work that considers the impact of multiple interferers and CSI uncertainty on beamforming in the context of underlay cognitive radio systems which allows both PUs and SUs to operate in a two-way relaying mode. The existing relay nodes can also obtain the chance of spectrum access in exchange for providing an amplify-and forward  cooperative service based on property-right spectrum sharing model. The main contribution of this work is to find the optimal beamforming vector which maximizes the QoS for both PU and SU transceivers in the above-mentioned system. The optimization problem is formulated as finding the beamform vector of the relay nodes that maximizes the QoS for PUs and SUs in the presence of unfriendly interferers with imperfect CSI. We consider the most general scenario with respect to uncertainty in interferers' CSI knowledge, in which no information is available about the distribution of such CSIs or its stochastic parameters. In the proposed model, we only consider a limited bound on the uncertainty of the interferers' CSI knowledge and design a robust solution that accounts for the worst-case scenario.

Here we study two cases, where in the first one a complete knowledge of CSI of the channels between the interferers and the PUs, the SUs, and the secondary relays is available. The feasibility of the pristine optimization problem is examined and closed form equations for the feasibility conditions are derived for this case. The feasibility condition leads us to provide an upper bound on the optimal PUs' SINR. Afterward, a solution for the SINR optimization problem is proposed. The bisection method is applied to obtain the optimal SINR which can be achieved by the PU and SU transceivers.
In second case , we consider the impact of imperfect knowledge of interferers' CSI on the designed system and calculate a robust solution for the SINR optimization problem. Another mathematical contribution of this work is to obtain the closed-form formulation of the worst-case scenario for each constraint instead of using linear matrix inequality (LMI) approaches with additional variables.

The rest of this paper is organized as follows: In Section \ref{sec:related}, an overview of some related works in the literature is presented. Section \ref{Sec:System} describes the system model. In Section \ref{Sec:PerfectCSI}, the SINR optimization problem is defined and solved with the assumption of perfect CSI knowledge for the unfriendly interferers. Subsequently, in Section \ref{Sec:ImperfectCSI}, we show how to obtain the solution for the SINR optimization problem if only an imperfect knowledge of interferers' CSI is available. Numerical results are provided in Section \ref{Sec:Simulation}; and Section \ref{Sec:Conclusion} draws the concluding remarks.

\section{Related Works} \label{sec:related}
In general, cooperative communication techniques have proven to significantly enhance the performance of wireless communication systems in terms of reducing the energy consumption, enhancing the transmission rate, and extending the connectivity, to only name a few \cite{Dai_Survey,Li_cooperative,Sami_Cooperative}. A growing body of literature has investigated different factors that play a key role in optimizing the performance of cooperative relaying systems including studying the impact of power allocation, relay selection, relaying modes and time allocation strategies among direct and cooperative communication \cite{afghah2013stochastic,Nam_Selection,Chen_2011}.
Furthermore, several joint optimization techniques have been proposed with the goal of improving the network performance when looking at the combined effect of these factors \cite{Liu_joint,Mo_DF,Mo_AF,Ng_Joint,Su_ICC}. The authors in \cite{Mo_DF} and \cite{Mo_AF} aimed at generalizing the common assumption of equal time allocation between the source and relay nodes and designed optimum joint power and time allocation mechanisms to minimize the outage probability when only the statistical knowledge of CSI is available.

Motivated by the results of cooperative relaying in wireless networks, the SUs have been deployed as relays in cognitive radio networks to enhance the QoS of PUs, particularly when the PUs experience a poor channel condition due to shadowing or sparse network coverage \cite{Simeone2008,CR-Afghah2013,Su_Active_2012}. Cooperative spectrum leasing solutions have recently  received a considerable attention in cognitive radio networks as they offer a coordination mechanism between the licensed and licensed users for dynamic spectrum access. In this methods, the secondary users can obtain the chance of spectrum access in exchange for providing cooperative services for the spectrum owners when they face poor channel conditions \cite{CR-Afghah2013,afghah2014cooperative,Simeone2008,Afghah_INFOCOM,Jayaweera}.
In \cite{Simeone2008}, a model for cooperative spectrum leasing among a primary user and a network of Ad-Hoc secondary users is presented in which the primary user can decide whether to lease a portion of its spectrum access time to the secondary users noting its channel quality. Moreover, a non-cooperative game theoretic model is defined to determine the optimum power allocation of the secondary users when they compete with one another to enhance their transmission rate over the assigned time for SUs' transmission. In \cite{Su_Active_2012}, the authors studied the cooperative spectrum leasing in heterogeneous Ad-Hoc networks and calculated the necessary condition on the channel quality between the primary user and the SUs to encourage primary users to participate in leasing. A cooperation protocol is proposed to maximize the transmission rate of secondary cognitive users for the given amount of spectrum released by the PU and their given power budget, where an equal time allocation is assigned for transmission of the primary and secondary users. Similar to any cooperative communication networks, the performance of the cooperative spectrum leasing techniques depend on several factors such as relay selection, deployed relaying methods, availability of global CSI, reliability of the secondary users, and presence of jamming or interference.

Different relaying strategies, including decode and forward (DF), compress and forward (CF), and amplify and forward (AF) are investigated in literature \cite{kim2008comparison}. The AF relaying mode has been widely utilized in practical applications due to its simplicity, as the relay nodes are only required to amplify and phase steer, i.e., beamform, the received signal and rebroadcast it.  Despite the DF and CF coding relaying techniques, where the relay nodes need to decode and re-encode the transmitter's message; in AF relaying mode, the relay nodes only amplify and forward the received signal. Therefore, AF involves lower complexity and are an appropriate relaying solution for cooperative spectrum sharing applications as selected in this paper since the SUs do not require to have the knowledge of PU's codebooks for relaying \cite{Truong,Su_Active_2012}.

While in cooperative spectrum leasing models, it is assumed that the secondary relays follow the agreement among the users to only transmit their messages in their allocated time slots \cite{CR-Afghah2013,afghah2014cooperative,Simeone2008,Jayaweera,Korenda,Namvar}, it is likely that these users will deviate from this agreement and cause harmful interference for the spectrum owners. Such undesired interference can be also caused by other unlicensed users in the proximity of the primary users working based on other spectrum sharing schemes such as spectrum sensing. In these conditions, the information about the channel conditions among the interferers and the PUs is often unavailable due to the lack of coordination among these users. This calls for new models to provide robust solutions to combat such undesired interference when minimum amount of information is accessible about the interferers. 
In general, three different approaches are known to handle the difficulties imposed by uncertainty in a data set, which in our case is imperfection in the interferers' CSI, \cite{ben1998robust}: 1) stochastic programming (SP), 2) robust mathematical programming (RMP), and 3) robust counterpart (RC) approach (worst-case scenario). The SP is limited to the problems where the uncertainty is stochastic in nature, i.~e.~ can be modeled as random variables. In this approach, the system is robustly designed in such a way that the average of network constraints including interference level or minimum QoS thresholds are in a desired range \cite{suraweera2010capacity,zhang2013outage,jamal2012performance,ozcan2016energy,Zarakovitis2016,Mo_DF,Mo_AF}. One may think of two practical drawbacks for this approach. First, we need to be able to identify the underlying probability distributions of uncertainty in the data set. The other vital pragmatic drawback of this approach is that it is very likely that the constraints on the average will be violated. Similarly, in the RMP approach, the violation of the constraints can occur but with a penalty in the objective  \cite{mulvey1995robust}. Hence, this cannot be a proper option in software defined radio (SDR) networks, where the constraints on PUs' QoS must be met accurately. Finally, RC approaches, first introduced by Ben-Tal and Nemrovski \cite{ben1998robust}, comply with the specified constraints in a problem by considering a worst-case scenario. To be more precise, a bounded region is assumed for the uncertainty on the variables that are not perfectly known, while the objective of RC method is to find a robust solution that is feasible over the whole uncertainty region. Commonly, a mathematical concept known as the  linear matrix inequality (LMI) along with defining new variables is used to solve an RC problem such as robust optimization problems in underlay SDR systems with uncertainty in CSI \cite{li2014robust,Tian2016,CRBidirectional-zaeri2016,wang2011robust}.

{\textit{Notations:}} A summary of the notation is provided in Table \ref{Table:notations}.
\begin{table}
  \centering
  \caption{Notation}\label{Table:notations}
  \begin{tabular}{|c|c|}
    \hline
    Notation & Description \\
    \hline \hline
    uppercase boldface letters & matrices \\
    \hline
    lowercase boldface letters &  vectors \\
    \hline
    $(.)^*$ & conjugate of the complex scalar \\
    \hline
    $(.)^T$ & transpose of a vector or matrix \\
    \hline
    $(.)^H$ & Hermitian (conjugate) transpose of a vector or matrix \\
    \hline
    $\|\ba\|$ & Euclidean norm ($\|.\|_2$) of the vector $\ba$ \\
    \hline
    $\lambda_{max}\{\bA\}$ & maximum eigenvalue of the matrix $\bA$\\
    \hline
    $\bA_{{i,j}}$ or $\ba_{{i}}$ & the ${\{i,j\}}^{\mbox{th}}$ or ${\{i\}}^{\mbox{th}}$ element of matrix $\bA$ or vector $\ba$\\
    \hline
    $\bA = \diag(\ba)$ & diagonal matrix with $\bA_{{i,i}} = \ba_{{i}}$ \\
    \hline
    $\ba = \diag(\bA)$ & vertical vector with $\ba_{{i}} = \bA_{{i,i}}$ \\
    \hline
    $\textsl{blkdiag}(\bA,\bB)$ & diagonal matrix with the elements of the vectors $\ba$ and $\bb$ in order\\
    \hline
    $\bA \bullet \bB$ & $\sum_i{\sum_j{\bA_{{i,j}}\bB_{{i,j}}}}$ \\
    \hline
    $\begin{array}{c}
        \\
        \bar{i}  \mbox{\;\;for\;} i = 1,2 \\
     \end{array}$ & $\begin{array}{c}
       \\
       \mbox{all except \;} i\\
     \end{array}$ \\
     \hline
    $\boldsymbol{\ell}_{N,j}$ & $N \times 1$ vector with $j$th element 1 and the rest 0\\
    \hline
  \end{tabular}
\end{table}

\section{System Model}\label{Sec:System}
\begin{figure}
\centerline{\resizebox{!}{6.5cm}{\includegraphics{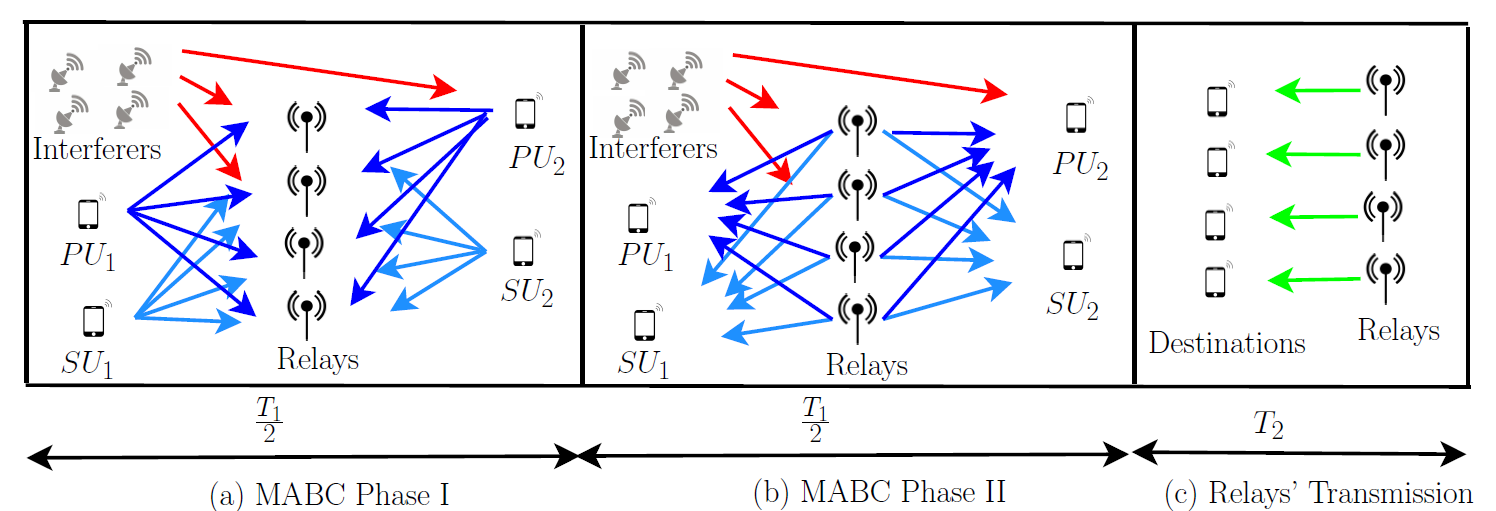}}}
 \caption{(a) MABC phase I in which both the PUs and SUs transmit their signals, simultaneously, and the relays receive these signals in presence of interferers. (b) MABC Phase II, in which the relays broadcast an amplified and phase shifted version of the signals they received in Phase I. (c) As a reward, the relays transmit their signal to their own destinations in a portion of this time-slot. $T_1$ refers to a portion of time slot allocated to transmission of the PUs and the SUs, while $T_2$ is the portion of the time slot allocated to the relays' transmission as an award for their cooperative services.}
 \label{fig:MABC}
 \end{figure}

In this section, the system and channel models considered in the paper are briefly outlined. A SDR network consists of two primary transceivers ($PU_1$ and $PU_2$) and two secondary transceivers ($SU_1$ and $SU_2$) is considered. The SUs are in a soft lease contract with the PUs, meaning that they obtained the permission to simultaneously access the PUs' spectrum based on an agreement \cite{sboui2015achievable}. Due to a low quality of the direct link between $PU_1$ and $PU_2$, the primary network is willing to employ several secondary relays for the sake of cooperative relaying advantages. An MABC two-way cooperative scenario is designed, where two PU transceivers, as well as two secondary transceivers, desire to exchange their signals with the help of $N_r$ relays, simultaneously. As a reward, the relays, called $R_1,R_2,\ldots,R_{N_r}$, will obtain access to the PU's spectrum during each communication cycle for a fixed portion of the time slot, as depicted in Fig.~ \ref{fig:MABC}. It is also assumed that the network is affected by $N_I$ unfriendly interferers $\{I_1,I_2, \ldots, I_{N_I}\}$ and that all wireless channels are reciprocal and frequency flat.

We assume that, in a given time-slot, the antennas can only transmit or receive a signal but not both at the same time, i.~e.~ all antennas operate in half-duplex mode. A centralized controller is considered to provide the perfect CSI of the SDR system as well as to calculate the optimum beamforming vector of the relays, an assumption which has been considered in similar reported works \cite{CRBidirectional-zaeri2016,CRBidirectional-Shahram2015,CRBidirectional-Zaeri2012,CRBidirectional-Veria2010}. The parameters of the system model are summarized in Table \ref{Table:Summaryparameters}.

\begin{table}
  \centering
  \caption{Summary of parameters}\label{Table:Summaryparameters}
  \begin{tabular}{|c|c|}
    \hline
    Parameters & Description \\
    \hline \hline
    $PU_i$ & i'th primary user \\
    \hline
    $SU_i$ & i'th Secondary users \\
    \hline
     $\bof_{P_i}$ & channel coefficient vectors between $PU_i$ and relays \\
    \hline
     $\bof_{S_i}$ & channel coefficient vectors between $SU_i$ and relays\\
    \hline
    $\bh_{P_i}$ & channel coefficient vectors between the interferers and $PU_i$ \\
    \hline
     $\bh_{S_i}$ & channel coefficient vectors between the interferers and $SU_i$ \\
    \hline
    $\bh_{I_l}$ & channel coefficient vectors between the interferers and l'th relay \\
    \hline
     $\br$ & received vector signal at relays \\
    \hline
     $\bw$ & beamformer vector\\
    \hline
    $\bt$ & transmitted signal by relays\\
    \hline
    $y_{P_i}$& received signal by $PU_i$\\
    \hline
    $y_{S_i}$& received signal by $SU_i$\\
    \hline
    $SINR_{P_i}$& SINR at $PU_i$\\
    \hline
    $SINR_{S_i}$& SINR at $SU_i$\\
    \hline
  \end{tabular}
\end{table}

We assume that the $N_r \times 1$ complex channel coefficient vectors $\bof_{P_i}$, $i = 1,2$, are defined as:
\begin{IEEEeqnarray}{rCl}\label{Def:fPiSr}
  \bof_{P_i} &=& [f_{P_iR_1}, f_{P_iR_2}, \ldots, f_{P_iR_{N_r}}]^T,
\end{IEEEeqnarray}
 where $f_{P_iR_j}$ is the instant reciprocal flat fading channel coefficient between $PU_i$ and the secondary relay $R_j$ for $i = 1,2$ and $j = 1,2,\ldots, N_r$. The $N_r \times 1$ complex channel coefficient vectors $\bof_{S_i}$, $i = 1,2$, are denoted by:
\begin{IEEEeqnarray}{rCl}\label{Def:fSiSr}
 \bof_{S_i} &=& [f_{S_iR_1}, f_{S_iR_2}, \ldots, f_{S_iR_{N_r}}]^T,
\end{IEEEeqnarray}
 where $f_{S_iR_j}$s, $i = 1,2$ and $j = 1,2,\ldots,N_r$ are the instant reciprocal flat fading channel coefficients between the transceiver $SU_i$ and the secondary relay $R_j$.

\begin{figure}
\centerline{\resizebox{!}{6.5cm}{\includegraphics{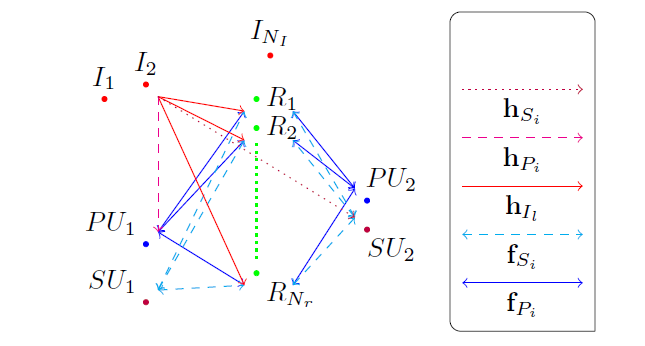}}}
\caption{Channel coefficient vectors between interferers and $SU_i$ ($\bh_{S_i}$), interferers and $PU_i$ ($\bh_{P_i}$), interferer $I_l$ and relays ($\bh_{I_l}$), $SU_i$ and relays($\bof_{S_i}$), and $PU_i$ and relays($\bof_{S_i}$).}
\label{fig:Channels}
 \end{figure}

Moreover, we assume that the $N_I \times 1$ complex channel coefficient vectors between the interferers and $PU_i$ and $SU_i$, $i=1,2$, are defined as:
 \begin{IEEEeqnarray}{rCl}\label{Def:hPiOrSi}
  \bh_{P_i} &=& [h_{P_i{I_1}},h_{P_i{I_2}},\ldots,h_{P_i{I_{N_I}}}]^T \IEEEyesnumber \IEEEyessubnumber \\
  \bh_{S_i} &=& [h_{S_i{I_1}},h_{S_i{I_2}},\ldots,h_{S_i{I_{N_I}}}]^T, \IEEEyessubnumber
\end{IEEEeqnarray}
where $h_{P_i{I_l}}$ and $h_{S_i{I_l}}$ denote the instant reciprocal flat fading channel coefficients between the interferer $I_l$, $l = 1,\ldots,N_I$, and $PU_i$ and $SU_i$, $i = 1,2$, respectively.
Also, the $N_r \times 1$ complex channel coefficient vectors between the secondary relays and the interferer $I_l$, $l = 1,2,\ldots,N_I$, are defined as:
\begin{IEEEeqnarray}{rCl}\label{Def:hSrIi}
  \bh_{I_l} &=& [h_{R_1I_l}, h_{R_2I_l}, \ldots,h_{R_{N_r}I_l}]^T,
\end{IEEEeqnarray}
 where, $h_{R_jI_l}$s, $j = 1,2,\ldots,N_r$ and $l = 1,2,\ldots,N_I$, are the instant reciprocal flat fading channel coefficients between the secondary relay $R_j$ and the interferer $I_l$. The channel coefficients in the system model are summarized in Fig.~\ref{fig:Channels}.

The complete knowledge of CSI between the interferers and the SDR users is not available and only  imperfect CSI estimations of channels between the PUs and the interferers, the SUs and  interferers, and the secondary relays and the  the interferers  denoted by $\hat{\bh}_{P_i}$, $\hat{\bh}_{S_i}$, $\hat{\bh}_{I_l}$, $i = 1,2$ and $l=1,2,\ldots,N_I$, respectively is provided by the centralized controller.  Mathematically, this assumption can be written as  \cite{CRBidirectional-zaeri2016}:
 \begin{IEEEeqnarray}{rCl}\label{Equ:ChannelIntBounded}
  {\bh_{P_i}} =& \hat{\bh}_{P_i} + \nabla\bh_{P_i} ,& \| \nabla\bh_{P_i} \| \leq \epsilon_{P_i}  \IEEEyesnumber  \IEEEyessubnumber\\
  {\bh_{S_i}} =& \hat{\bh}_{S_i} + \nabla\bh_{S_i} ,& \| \nabla\bh_{S_i} \| \leq \epsilon_{S_i}  \IEEEyessubnumber\\
  {\bh_{I_l}} =& \hat{\bh_{I_l}} + \nabla\bh_{I_l} ,& \| \nabla\bh_{I_l} \| \leq \epsilon_l  \IEEEyessubnumber
\end{IEEEeqnarray}
where $\nabla\bh_{P_i}$, $\nabla\bh_{S_i}$ and $\nabla\bh_{I_l}$ are the bounded uncertainty CSI vectors and $\epsilon_{P_i}$, $\epsilon_{S_i}$, and $\epsilon_l$ denote the maximum value of CSI estimation error. The key advantage of this model is that it does not rely on the knowledge of distribution of the estimation errors rather it only requires the maximum value of these errors \cite{CRBidirectional-zaeri2016}.

In MABC protocol, the messages are exchanged in two time slots, where in the first one, the primary and secondary transceivers send their messages, $x_{P_i}$ and $x_{S_i}$, $i = 1,2$, simultaneously. The relays receive a linear combination of all primary and secondary transmitted signals as well as the external interference signals $x_{I_l}^{(1)}$, $l = 1,2,\ldots,N_I$. Each relay rebroadcasts a weighted version of the received signal (AF relaying) in the second time-slot. Each transceiver receives a linear combination of its own signal, the other primary or secondary transceivers and the external interference signals $x_{I_l}^{(2)}$, $l = 1,2,\ldots,N_I$, in this time-slot. Without loss of generality, we assume that {$\mathbf{E}\{|x_{P_i}|^2\} = \mathbf{E}\{|x_{S_i}|^2\} = \mathbf{E}\{|x_{I_l}^{(i)}|^2\} = 1$} for all $i = 1,2$ and $l = 1,\ldots,N_I$. Also, we assume that all messages from different sources or in different time-slots are independent. Either primary or secondary transceivers may extract the desired signal considering the full knowledge of the self-interference portion of the signal.

The $N_r \times 1$ vector of the received signal at the secondary relay network, in the first time-slot, can be written as:
 \begin{IEEEeqnarray}{rCl}\label{Equ:SignalRelayRec}
  \br &=& \sum_{i=1}^{2}{\sqrt{P_{P_i}}\bof_{P_i}x_{P_i}} + \sum_{i=1}^{2}{\sqrt{P_{S_i}}\bof_{S_iR
  }x_{S_i}} + \sum_{l=1}^{N_I}{\sqrt{P_{I_l}}\bh_{I_l}x_{I_l}^{(1)}} + \boldsymbol{\nu},
\end{IEEEeqnarray}
 where $P_{P_i}$, $P_{S_i}$ and $P_{I_l}$, $i = 1,2$ and $l = 1,2,\ldots,N_I$ denote the transmit power of primary transceivers, secondary transceivers and interferers, respectively. The $N_r \times 1$ vector $ \boldsymbol{\nu} \sim \mathcal{N}_{\mathcal{C}}(\textbf{0}, \sigma^2 \bI_{N_r})$ represents the white Gaussian noise at the relays.
Each relay multiplies its received signal by a complex number $w^*_j$, $j = 1,2,\ldots,N_r$ and rebroadcasts it in the second time-slot. By defining the $N_r \times 1$ beamforming vector $\bw = [w_1,w_2,\ldots,w_{N_r}]^T$, the $N_r \times 1$ transmitted vector at the relays can be written as:
 \begin{IEEEeqnarray}{rCl}\label{Equ:SignalRelayTra}
  \bt &=& \bW^H \br \IEEEyesnumber\\
  &=& \sum_{i=1}^{2}{\sqrt{P_{P_i}}\bW^H\bof_{P_i} x_{P_i}} + \sum_{i=1}^{2}{\sqrt{P_{S_i}}\bW^H\bof_{S_i} x_{S_i}} + \sum_{l=1}^{N_I}{\sqrt{P_{I_l}}\bW^H \bh_{I_l}  x_{I_l}^{(1)}} + \bW^H \boldsymbol{\nu}. \IEEEnonumber
\end{IEEEeqnarray}
The individual power consumption at each relay can be written as:
\begin{IEEEeqnarray}{rCl}\label{Equ:Pi_individual}
  P_{r_j} &=& \mathbf{E}\{ |\bt_{\{j\}}|^2 \} = (\bW^H \mathbf{E}\{ \br \br^H \}\bW)_{\{j,j\}} \IEEEnonumber \\
  &=&  (\bW^H\left\{ \sum_{i=1}^{2}{{P_{P_j}}\bof_{P_iR}\bof_{P_iR}^H} + \sum_{i=1}^{2}{{P_{S_i}}\bof_{S_iR}\bof_{S_iR}^H} +
   \sum_{l=1}^{N_I}{{P_{I_l}}\bh_{I_l}\bh_{I_l}^H} + \sigma^2 \bI_{N_r}\right\}\bW)_{\{j,j\}}, \IEEEyesnumber
\end{IEEEeqnarray}
for $j = 1,\ldots,N_r$. After some mathematical manipulation and by using the fact that $\bA\bb = \bB\ba$ if $\ba=diag(\bA)$ and $\bb=diag(\bB)$ for arbitrary same-size vectors $\ba$ and $\bb$, (\ref{Equ:Pi_individual}) is simplified as:
\begin{IEEEeqnarray}{rCl}\label{Equ:Pi_individual_simple}
  P_{r_j} = \xi_{r_j} |w_j|^2,
\end{IEEEeqnarray}
where
\begin{IEEEeqnarray}{rCl}\label{Def:xi}
    \xi_{r_j} &=& \sum_{i=1}^{2}{{P_{P_i}}{|(\bof_{P_iR})_{\{j\}}|^2}} + \sum_{i=1}^{2}{{P_{S_i}}|(\bof_{S_iR})_{\{j\}}|^2} + \sum_{l=1}^{N_I}{{P_{I_l}}|(\bh_{I_l})_{\{j\}}|^2} + \sigma^2, \IEEEyesnumber
\end{IEEEeqnarray}
for $j = 1,\ldots,N_r$. The received signal at $PU_i$, $i=1,2$, in the second time-slot can be written as:
 \begin{IEEEeqnarray}{rCl}\label{Equ:SignalPrimaryRec}
  y_{P_i} &=& \sum_{k=1}^{2}{\sqrt{P_{P_k}}\bw^H\bF_{P_i}\bof_{P_k} x_{P_k}}
  + \sum_{k=1}^{2}{\sqrt{P_{S_k}} \bw^H \bF_{P_i} \bof_{S_k} x_{S_k}} \IEEEnonumber \\
      &+& \sum_{l=1}^{N_I}{\sqrt{P_{I_l}}\bw^H\bF_{P_i}\bh_{I_l} x_{I_l}^{(1)}} +  \sum_{l=1}^{N_I}{\sqrt{P_{I_l}} h_{P_i{I_l}}x_{I_l}^{(2)}}
      + \bw^H \bF_{P_i}\boldsymbol{\nu} + \nu_{P_i},\IEEEyesnumber
\end{IEEEeqnarray}
where $\nu_{P_i} \sim \mathcal{N}_{\mathcal{C}}(0, \sigma^2)$ is the white Gaussian noise at the $PU_i$,  $\bw$ denotes the beamforming vector. Notation $\bF$ represents a diagonal matrix with $\bF_{{i,i}}=\bf_{{i}}$ as previously defined in Table \ref{Table:notations}. 
The received signal at $SU_i$, $i=1,2$, in the second time-slot can be written as: \begin{IEEEeqnarray}{rCl}\label{Equ:SignalSecondaryRec}
  y_{S_i} &=& \sum_{k=1}^{2}{\sqrt{P_{P_k}}\bw^H \bF_{S_i} \bof_{P_k} x_{P_k}}
  + \sum_{k=1}^{2}{\sqrt{P_{S_k}}\bw^H \bF_{S_i} \bof_{S_k}  x_{S_k}} \IEEEnonumber \\
      &+& \sum_{l=1}^{N_I}{\sqrt{P_{I_l}}\bw^H \bF_{S_i} \bh_{I_l} x_{I_l}^{(1)}} +  \sum_{l=1}^{N_I}{\sqrt{P_{I_l}} h_{S_i{I_l}}x_{I_l}^{(2)}}
      + \bw^H \bF_{S_i} \boldsymbol{\nu} + \nu_{S_i},\IEEEyesnumber
\end{IEEEeqnarray}
where $\nu_{S_i} \sim \mathcal{N}_{\mathcal{C}}(0, \sigma^2)$ is the white Gaussian noise at the $SU_i$.
It is assumed that each primary and secondary transceiver can compute and subtract the self-interference part of the received signal. Therefore, the residual received signals at the primary transceiver can be written as:{\begin{IEEEeqnarray}{rCl}\label{Equ:SignalPrimaryRecRes}
  \tilde{y}_{P_j} &=& \underbrace{\sqrt{P_{P_{\bar{j}}}}\bw^H\bF_{P_j}\bof_{P_{\bar{j}}R} x_{P_{\bar{j}}}}_{\mbox{Desired Signal}}
  + \sum_{i=1}^{2}{\sqrt{P_{S_i}}\bw^H \bF_{P_j} \bof_{S_i}  x_{S_i}} \IEEEnonumber\\
      &+& \sum_{l=1}^{N_I}{\sqrt{P_{I_l}}\bw^H \bF_{P_j} \bh_{I_l} x_{I_l}^{(1)}} +  \sum_{l=1}^{N_I}{\sqrt{P_{I_l}} h_{P_j{I_l}}x_{I_l}^{(2)}}
      + \bw^H \bF_{P_j} \boldsymbol{\nu} + \nu_{P_j}, \;\;\; j = 1,2. \IEEEyesnumber
\end{IEEEeqnarray}}
Also the residual received signals at the secondary transceivers can be written as:
\begin{IEEEeqnarray}{rCl}\label{Equ:SignalSecondaryRecRes}
  \tilde{y}_{S_j} &=& \sum_{i=1}^{2}{\sqrt{P_{P_i}}\bw^H \bF_{S_j}\bof_{P_i}  x_{P_i}}
  + \underbrace{\sqrt{P_{S_{\bar{j}}}}\bw^H \bF_{S_j} \bof_{S_{\bar{j}}R}  x_{S_{\bar{j}}}}_{\mbox{Desired Signal}} \IEEEnonumber \\
      &+& \sum_{l=1}^{N_I}{\sqrt{P_{I_l}}\bw^H \bF_{S_j}\bh_{I_l}  x_{I_l}^{(1)}} +  \sum_{l=1}^{N_I}{\sqrt{P_{I_l}} h_{S_j{I_l}}x_{I_l}^{(2)}}
      + \bw^H\bF_{S_j}\boldsymbol{\nu} + \nu_{S_j}, \;\;\; j = 1,2. \IEEEyesnumber
\end{IEEEeqnarray}
Let us define
\begin{IEEEeqnarray}{rCl}\label{Def:k}
  \bk_{P_iP_j} = \bF_{P_i}\bof_{P_j} \qquad , \qquad
  \bk_{S_iS_j} = \bF_{S_i}\bof_{S_j} \qquad \mbox{and} \qquad
  \bk_{S_iP_j} = \bF_{S_i}\bof_{P_j},\IEEEyesnumber
\end{IEEEeqnarray}
for $i,j = 1,2$,
\begin{IEEEeqnarray}{rCl}\label{Def:Q}
  \bQ_{P_j} &=& \sum_{i=1}^{2}{P_{S_i}\bk_{S_iP_j}\bk_{S_iP_j}^H}+\sigma^2 \bF_{P_j} \bF_{P_j}^H
  + \sum_{l=1}^{N_I}{P_{I_l}{\bF_{P_j}\bh_{I_l}\bh_{I_l}^H\bF_{P_j}^H}}, \;\;\; j = 1,2 \IEEEyesnumber \IEEEyessubnumber\\
  \bQ_{S_j} &=& \sum_{i=1}^{2}{P_{P_i}\bk_{S_jP_i}\bk_{S_jP_i}^H}+\sigma^2 \bF_{S_j} \bF_{S_j}^H
  + \sum_{l=1}^{N_I}{P_{I_l}{\bF_{S_j}\bh_{I_l}\bh_{I_l}^H\bF_{S_j}^H}}, \;\;\; j=1,2.\IEEEyessubnumber
\end{IEEEeqnarray}
and
\begin{IEEEeqnarray}{rCl}\label{Def:zeta}
  \zeta_{P_i} = \sum_{l=1}^{N_I}{P_{I_l}|h_{P_i{I_l}}|^2} \qquad , \qquad
  \zeta_{S_i} = \sum_{l=1}^{N_I}{P_{I_l}|h_{S_i{I_l}}|^2}, \;\;\; i=1,2.\IEEEyesnumber
\end{IEEEeqnarray}
By using the above definitions and the residual signals (\ref{Equ:SignalPrimaryRecRes}) and (\ref{Equ:SignalSecondaryRecRes}), the SINRs at the primary and secondary transceivers can be written as:
\begin{IEEEeqnarray}{rCl}{\label{Equ:SINRPerfect}}
   SINR_{P_i} &=& \frac{P_{P_{\bar{i}}}\bw^H\bk_{P_1P_2}\bk_{P_1P_2}^H \bw}
  { \bw^H\bQ_{P_i}\bw+\zeta_{P_i}+\sigma^2}, \;\;\; i =1,2 \IEEEyesnumber \IEEEyessubnumber\\
  SINR_{S_i} &=& \frac{P_{S_{\bar{i}}}\bw^H\bk_{S_1S_2}\bk_{S_1S_2}^H\bw}
  { \bw^H\bQ_{S_i}\bw+\zeta_{S_i}+\sigma^2}, \;\;\; i =1,2. \IEEEyessubnumber
\end{IEEEeqnarray}
In the next section, the SINR optimization problem is discussed.
\section{SINR's Optimization with perfect CSI}\label{Sec:PerfectCSI}
The goal of this section is to find an optimal weight vector, $\bw$ such that the SINRs in the primary and secondary network is maximized. This optimization problem can be represented as a Max-Min problem with the sense of maximizing the minimum value among primary and secondary's SINRs. Despite the majority of reported works, in which the optimal beamforming vector is found to guarantee a minimum QoS for only the PUs, here we provide a solution to assure the required QoS for both PUs and SUs. In fact, the SUs have already obtained access to the spectrum by reason of their soft-lease and deserve to have a minimum QoS. However, considering the priority of the PUs as the spectrum owners, a design parameter $ \mu \geq 1$ is defined as an expected ratio between SINRs for the PUs and SUs. Another assumption in this optimization problem is that the available individual power at the relays is limited. Therefore, the SINR maximization problem can be written as:

\begin{IEEEeqnarray}{rCl}\label{Equ:OptimizationMaxMin}
  \max_{\bw} \mbox{\;}& \min{\left\{\{SINR_{P_i}\}_{i = 1,2},\mu \{SINR_{S_i}\}_{i = 1,2}\right\}} & \IEEEeqnarraynumspace \IEEEyesnumber \\
 & \mbox{Subject To:\;} ~~ P_{r_j} \leq P_l^{max},& \;\;\; l = 1,\ldots,N_r \IEEEyessubnumber.
\end{IEEEeqnarray}
By defining an auxiliary real variable $\gamma > 0$, the Max-Min problem (\ref{Equ:OptimizationMaxMin}) can be rewritten as:
 \begin{IEEEeqnarray}{rCl}\label{Equ:Optimizationwgamma}
  \max_{\bw, \gamma>0} & \gamma & \IEEEyesnumber\\
  \mbox{Subject To:}& SINR_{P_i} \geq \gamma,  & \mbox{\;\;\;} i = 1,2\IEEEyessubnumber \\
  &\mu SINR_{S_i} \geq \gamma, & \mbox{\;\;\;} i = 1,2\IEEEyessubnumber \\
   & P_{r_j} \leq P_j^{max}, & \mbox{\;\;\;} j = 1,\ldots,N_r \IEEEyessubnumber.
\end{IEEEeqnarray}
 By using (\ref{Equ:Pi_individual_simple}) and (\ref{Equ:SINRPerfect}), the optimization problem (\ref{Equ:Optimizationwgamma}) can be rewritten as:
\begin{IEEEeqnarray}{rCl}\label{Equ:OptimizationwgammaExpand}
  \max_{\bw, \gamma>0} & \gamma& \IEEEyesnumber\\
  \mbox{Subject To:}&\frac{P_{P_{\bar{i}}}\bw^H\bk_{P_1P_2}\bk_{P_1P_2}^H \bw}
  { \bw^H\bQ_{P_i}\bw+\zeta_{P_i}+\sigma^2} \geq \gamma, & \mbox{\;\;\;} i = 1,2 \IEEEyessubnumber \IEEEeqnarraynumspace\\
  &\mu \frac{P_{S_{\bar{i}}}\bw^H\bk_{S_1S_2}\bk_{S_1S_2}^H\bw}
  { \bw^H\bQ_{S_i}\bw+\zeta_{S_i}+\sigma^2} \geq \gamma, & \mbox{\;\;\;} i = 1,2 \IEEEyessubnumber \IEEEeqnarraynumspace\\
   & \xi_{r_j} |{\boldsymbol{\ell}_{N_r,j}}^T \bw|^2 \leq P_j^{max}, & \mbox{\;\;\;} j = 1,\ldots,N_r \IEEEyessubnumber.
   \IEEEeqnarraynumspace
\end{IEEEeqnarray}
In the following subsection we investigate the feasibility condition(s) of the optimization problem (\ref{Equ:OptimizationwgammaExpand}).
\subsection{Feasibility Condition}
\begin{lemma}\label{Lemma:OneEigen}
Let $\bDelta$ denote a Positive Definite (PD) matrix, $\ba$ represent a vector of the same length as size of $\bDelta$, t and c are positive scalars and $\bx$ be the vector variable with the same length as $\ba$, then $\ba^H \bDelta^{-1}\ba - t \geq 0$ is a feasibility condition for the following constraint:
\begin{IEEEeqnarray}{rCl}\label{Equ:Lemma1Const}
  \frac{\bx^H\ba\ba^H\bx}{\bx^H\bDelta\bx + c} \geq t.
\end{IEEEeqnarray}
\end{lemma}

\begin{proof}
The constraint (\ref{Equ:Lemma1Const}) can be rearranged and be written as:
\begin{IEEEeqnarray}{rCl}\label{Equ:Lemma1ConstV1}
 &&\bx^H(\ba\ba^H - t\bDelta)\bx \geq ct
\end{IEEEeqnarray}
An optimization becomes infeasible, i.~e.~ no feasible point for $\bx$ were found, if the core matrix $\ba\ba^H - t\bDelta$ was negative semi-definite. Also, if this matrix was not negative semi-definite, a vector $\bx_p$ exists such a way that $\bx_p^H(\ba\ba^H - t\bDelta)\bx_p > 0$, thus one may scale up $\bx_p$ in a way that the inequality (\ref{Equ:Lemma1ConstV1}) is satisfied. Therefore, (\ref{Equ:Lemma1ConstV1}) is feasible if and only if the matrix $\ba\ba^H - t\bDelta$ was not negative semi-definite.
As a result of the above discussion, the infeasibility condition can be written as:
\begin{IEEEeqnarray}{rCl}\label{Equ:Lemma1ConstNSD}
  &&\ba\ba^H - t\bDelta \preceq 0.
\end{IEEEeqnarray}
Since the matrix $\bDelta$ is positive definite, the constraint (\ref{Equ:Lemma1ConstNSD}) is equivalent to $\bDelta^{\frac{1}{2}}(\bDelta^{\frac{-1}{2}}\ba\ba^H\bDelta^{\frac{-1}{2}} - t\bI)\bDelta^{\frac{1}{2}} \preceq 0$. Also, the matrix $\bDelta^{\frac{1}{2}}$ is positive definite, we can rewrite the above mentioned condition as $\bDelta^{\frac{-1}{2}}\ba\ba^H\bDelta^{\frac{-1}{2}} - t\bI \preceq 0$, or equivalently $\lambda_{max}\{\bDelta^{\frac{-1}{2}}\ba\ba^H\bDelta^{\frac{-1}{2}} - t\bI \} > 0$. The matrix $\bDelta^{\frac{-1}{2}}\ba\ba^H\bDelta^{\frac{-1}{2}}$ is a rank one matrix. Hence, for $t > 0$, all of the eigenvalues of the matrix $\bDelta^{\frac{-1}{2}}\ba\ba^H\bDelta^{\frac{-1}{2}} - t\bI$ are equal to zero except $\ba^H\bDelta^{-1}\ba - t$. Therefore, the feasibility condition can be summarized as $\ba^H\bDelta^{-1}\ba - t > 0$.
\end{proof}

Using Lemma \ref{Lemma:OneEigen}, the feasibility conditions for the individual constraints in optimization problem (\ref{Equ:OptimizationwgammaExpand}) can be written as:
\begin{IEEEeqnarray}{rCl}\label{Equ:feasiblecon}
  &P_{P_{\bar{i}}}\bk_{P_1P_2}^H {\bQ_{P_i}}^{-1}
  \bk_{P_1P_2} - \gamma > 0,& \;\;\; i = 1,2 \IEEEyesnumber \IEEEyessubnumber\\
  &\mu P_{S_{\bar{i}}}\bk_{S_1S_2}^H {\bQ_{S_i}}^{-1}
  \bk_{S_1S_2} - \gamma > 0,& \;\;\; i = 1,2. \IEEEyessubnumber
\end{IEEEeqnarray}
Therefore, by using the feasibility conditions (\ref{Equ:feasiblecon}), an upper bound condition for $\gamma$ can be written as:
\begin{IEEEeqnarray}{rCl}\label{Equ:Uppergamma}
   \gamma^{up} &=& \min{\left\{\{P_{P_{\bar{i}}}\bk_{P_1P_2}^H {\bQ_{P_i}}^{-1}
  \bk_{P_1P_2}\}_{i=1,2 }
  ,  \mu \{P_{S_{\bar{i}}}\bk_{S_1S_2}^H {\bQ_{S_i}}^{-1}
  \bk_{S_1S_2}\}_{ i=1,2}\right\}}. \IEEEyesnumber
\end{IEEEeqnarray}
The upper bound value (\ref{Equ:Uppergamma}) of $\gamma$ does not guarantee the feasibility of the optimization problem (\ref{Equ:OptimizationwgammaExpand}), since it was obtained from individual constraints. In other word, the union of the feasibility regions associated with each constraint in (\ref{Equ:OptimizationwgammaExpand}) may provide additional limitation on $\gamma$ value. Therefore, we provide a method to find the optimal solution of $\gamma$ in the next subsection.
\subsection{Optimal SINR Solution}\label{Subsec:BisectionPerfect}
In this subsection, a feasibility check bisection method is used to find the optimal value of $\gamma$. In this method, the optimization problem (\ref{Equ:OptimizationwgammaExpand}) will turn into the following feasibility check problem for a given value of $\gamma$:
\begin{IEEEeqnarray}{rCl}\label{Equ:Findw}
  & \mbox{Find} \;\;\; {\bw} \IEEEyesnumber \\
  &\frac{P_{P_{\bar{i}}}\bw^H\bk_{P_1P_2}\bk_{P_1P_2}^H \bw}
  { \bw^H\bQ_{P_i}\bw+\zeta_{P_i}+\sigma^2} \geq \gamma, &\mbox{\;\;\;} i =1,2 \IEEEyessubnumber \\
  &\mu \frac{P_{S_{\bar{i}}}\bw^H\bk_{S_1S_2}\bk_{S_1S_2}^H\bw}
  { \bw^H\bQ_{S_i}\bw+\zeta_{S_i}+\sigma^2} \geq \gamma, & \mbox{\;\;\;} i =1,2  \IEEEyessubnumber\\
   & \xi_{r_j} |{\boldsymbol{\ell}_{N_r,j}}^T \bw|^2 \leq P_j^{max}, \mbox{\;\;\;} & j = 1,\ldots,N_r. \IEEEyessubnumber
\end{IEEEeqnarray}
The idea is to find the optimal value of $\gamma$, i.e., $\gamma^{opt}$, we do not need to calculate the optimum vector $\bw$. To do so, we start from an initial interval $(\gamma_0^{low},\gamma_0^{up})$ of $\gamma^{opt}$ where $\gamma_0^{up}$ is the initial upper bound of $\gamma^{opt}$, which is derived in (\ref{Equ:Uppergamma}) and $\gamma_0^{low}$ is the initial lower bound of $\gamma^{opt}$, which is zero. Then, at step $n$, by choosing $\gamma = \frac{1}{2}(\gamma_{n-1}^{low}+\gamma_{n-1}^{up})$ and checking the feasibility of problem (\ref{Equ:Findw}), the solution interval of $\gamma^{opt}$ will be updated as:
\begin{IEEEeqnarray}{rCl}\label{Equ:bisectionStep}
  (\gamma_n^{low},\gamma_n^{up}) &=& \left\{\begin{array}{cc}
                                            (\frac{1}{2}(\gamma_{n-1}^{low}+\gamma_{n-1}^{up}),\gamma_{n-1}^{up}) &  \mbox{, if (\ref{Equ:Findw}) feasible}\\
                                            &\\
                                            (\gamma_{n-1}^{low},\frac{1}{2}(\gamma_{n-1}^{low}+\gamma_{n-1}^{up})) & \mbox{, otherwise.}
                                          \end{array} \right. \IEEEeqnarraynumspace
\end{IEEEeqnarray}
This bisection method will be continued until a small enough range of $\epsilon$ for the solution interval of $\gamma^{opt}$ is achieved. It is worth mentioning that the bisection method increases the complexity order of our method by a factor of $\log_2\left(\frac{\gamma^{up}}{\epsilon}\right)$.
In order to solve the feasibility check problem (\ref{Equ:Findw}), we rearrange the constraints in a quadratic format:
\begin{IEEEeqnarray}{rCl}\label{Equ:FindwQuad}
  &\mbox{Find} \;\;\; {\bw} \IEEEyesnumber \\
  &\bw^H (P_{P_{\bar{i}}} \bk_{P_1P_2}\bk_{P_1P_2}^H - \gamma \bQ_{P_i}) \bw \geq \gamma(\zeta_{P_i}+\sigma^2), & \mbox{\;\;\;} i =1,2 \IEEEyessubnumber \\
  & \bw^H (\mu P_{S_{\bar{i}}}\bk_{S_1S_2}\bk_{S_1S_2}^H - \gamma \bQ_{S_i})\bw \geq  \gamma(\zeta_{S_i}+\sigma^2),& \mbox{\;\;\;} i =1,2 \IEEEyessubnumber \\
   & \xi_{r_j} |{\boldsymbol{\ell}_{N_r,j}}^T \bw|^2 \leq P_j^{max}, & \mbox{\;\;\;} j = 1,\ldots,N_r \IEEEyessubnumber.
\end{IEEEeqnarray}
It is observed that if $\bw$ is in the feasible region of (\ref{Equ:FindwQuad}), then for any arbitrary real number $\theta$, the vector $\tilde{\bw} = e^{j\theta} \bw$ is also in the feasible region. Therefore, without loss of generality, we can assume that $\bk_{P_1P_2}^H \bw$ (or $\bk_{S_1S_2}^H \bw$) is a non-negative real number. By applying this assumption, the constraints (\ref{Equ:FindwQuad}a) are turned to:
\begin{IEEEeqnarray}{rCl}\label{Equ:SOCPconst}
  &\sqrt{\frac{P_{P_{\bar{i}}}}{\gamma}}{\bk_{P_1P_2}}^H \bw
  \geq \sqrt{\bw^H \bQ_{P_i} \bw +\zeta_{P_i}+\sigma^2}, & \mbox{\;\;\;} i =1,2 \IEEEyesnumber \IEEEyessubnumber\\
  & \Re\{\bk_{P_1P_2}^H \bw\} \geq 0 \qquad , \qquad
   \Im\{\bk_{P_1P_2}^H \bw\} = 0 & \IEEEyessubnumber,
\end{IEEEeqnarray}
where (\ref{Equ:SOCPconst}a) constraints are complex second order (Lorentz) cone (SOC) \cite{boyd2004convex}. However, as an indirect conclusion from Lemma \ref{Lemma:OneEigen}, the matrices $\mu P_{S_{\bar{i}}}\bk_{S_1S_2}\bk_{S_1S_2}^H - \gamma \bQ_{S_i}$, $i = 1,2$, are not positive definite and therefore the constraints (\ref{Equ:FindwQuad}) do not represent convex regions. In order to make these constraints set convex, we define an auxiliary matrix variable $\bOmega = \bw \bw^H$.
 In a quadratic optimization problem, we are allowed to use the matrix variable $\bOmega$ by adding the following counterpart constraints \cite{palomar2010convex}:
\begin{IEEEeqnarray}{rCl}\label{Equ:OmegaPSD}
  \bOmega - \bw \bw^H  \succeq 0
\end{IEEEeqnarray}
and $rank(\bOmega) = 1$.
The positive semidefinite (PSD) condition (\ref{Equ:OmegaPSD}) represents a convex region for variables $\bOmega$ and $\bw$ \cite{boyd2004convex}. Although, the rank one condition does not represent a convex region; however, one may ignore the rank constraint of $\Omega$ and solve the relaxed optimization problem. It is shown that if a relaxed problem (non ranked restricted) which is linear with respect to $\bOmega$, was feasible then the rank-one restricted version of that problem is also feasible \cite{palomar2010convex}. Therefore, we remove the rank constraint from our optimization problem.

By using the matrix variable $\bOmega$, equation (\ref{Equ:FindwQuad}b) can be converted to:
\begin{IEEEeqnarray}{rCl}\label{Equ:QOPconstOmega}
&\gamma \bw^H \bQ_{S_i}\bw - \mu P_{S_{\bar{i}}}\bk_{S_1S_2}\bk_{S_1S_2}^H \bullet \bOmega + \gamma(\zeta_{S_i}+\sigma^2) \leq 0 & \mbox{\;\;\;} i =1,2  \IEEEyesnumber \IEEEyessubnumber \\
&\Re\{(\bk_{S_1S_2}\bk_{S_1S_2}^H) \bullet \bOmega \} \geq 0  \qquad , \qquad \Im\{(\bk_{S_1S_2}\bk_{S_1S_2}^H) \bullet \bOmega \} = 0 & \IEEEyessubnumber
\end{IEEEeqnarray}
for $j=1,2$.  In order to reduce the computational complexity order of the problem, the following lemma can be used to convert the quadratic region (\ref{Equ:QOPconstOmega}) into a SOC region.
\begin{lemma}\label{Lemma:QOP2SOCP}
For a vector $\ba$ and positive real numbers $\alpha \geq 0$ and $\beta \geq 0$, the inequality $\ba^H \ba \leq \alpha \beta$ holds if and only if the inequality  $\left\|\left[\begin{array}{c}
            \alpha - \beta \\
            2\ba
          \end{array}\right]\right\| \leq \alpha + \beta$ holds \cite{palomar2010convex}.
\end{lemma}
\begin{proof}
 The lemma will be simply proved by noting that $\left\|\left[\begin{array}{c}
     \alpha - \beta \\
            2\ba
          \end{array}\right]\right\|^2 = (\alpha - \beta)^2+4 \|\ba\|^2$.
\end{proof}
By applying Lemma \ref{Lemma:QOP2SOCP} for the values of $\alpha = 1$, $\beta = \frac{\mu}{\gamma} P_{S_{\bar{i}}}\bk_{S_1S_2}\bk_{S_1S_2}^H \bullet \bOmega - \zeta_{S_i}-\sigma^2$ and $\ba = {\bQ_{S_i}}^{\frac{1}{2}}\bw$, the constraints (\ref{Equ:QOPconstOmega}) will turned into the following SOC constraints:
\begin{IEEEeqnarray}{rCl}\label{Equ:SOCconstOmega}
  &\left\|\left[\begin{array}{c}
             1 - \frac{\mu P_{S_{\bar{i}}}}{\gamma} \bk_{S_1S_2}\bk_{S_1S_2}^H \bullet \bOmega +\zeta_{S_i}+\sigma^2 \\
            2{\bQ_{S_i}}^{\frac{1}{2}}\bw
          \end{array}\right]\right\| \leq 1 + \frac{\mu P_{S_{\bar{i}}}}{\gamma} \bk_{S_1S_2}\bk_{S_1S_2}^H \bullet \bOmega - \zeta_{S_i}-\sigma^2,& \;\;\; i = 1,2\IEEEyesnumber \IEEEyessubnumber\\
  &\Re\{(\bk_{S_1S_2}\bk_{S_1S_2}^H) \bullet \bOmega \} \geq 0  \qquad , \qquad
  \Im\{(\bk_{S_1S_2}\bk_{S_1S_2}^H) \bullet \bOmega \} = 0.  \IEEEyessubnumber
\end{IEEEeqnarray}
By using (\ref{Equ:SOCPconst}), (\ref{Equ:OmegaPSD}) and (\ref{Equ:SOCconstOmega}), the relaxed version of the problem (\ref{Equ:FindwQuad}) can be written as:
\begin{IEEEeqnarray}{rCl}\label{Equ:FindwSOCP}
  & \mbox{Find} \;\;\; {\bw,\bOmega} \IEEEyesnumber \IEEEnosubnumber\\
   &\sqrt{\gamma \bw^H \bQ_{P_i} \bw + \zeta_i^p+\sigma^2} \leq \sqrt{\frac{P_{P_{\bar{i}}}}{\gamma}}\bk_{P_1P_2}^H \bw,  & \mbox{\;\;\;} i =1,2 \IEEEnonumber  \\
   & \Re\{\bk_{P_1P_2}^H \bw\} \geq 0  \qquad
   \Im\{\bk_{P_1P_2}^H \bw\} = 0  & \IEEEnonumber \\
  &\left\|\left[\begin{array}{c}
             1 - \frac{\mu}{\gamma} P_{S_{\bar{i}}}\bk_{S_1S_2}\bk_{S_1S_2}^H \bullet \bOmega +\zeta_{S_i}+\sigma^2 \\
            2{\bQ_{S_i}}^{\frac{1}{2}}\bw
          \end{array}\right]\right\|
  \leq 1 + \frac{\mu}{\gamma} P_{S_{\bar{i}}}\bk_{S_1S_2}\bk_{S_1S_2}^H \bullet \bOmega - \zeta_{S_i} -\sigma^2, & \mbox{\;\;\;} i =1,2 \IEEEnonumber \\
    &\Re\{(\bk_{S_1S_2}\bk_{S_1S_2}^H) \bullet \bOmega \} \geq 0  \qquad ,\qquad
  \Im\{(\bk_{S_1S_2}\bk_{S_1S_2}^H) \bullet \bOmega \} = 0 & \IEEEnonumber \\
  & \sqrt{\xi_{r_j}}|{\boldsymbol{\ell}_{N_r,j}}^T \bw| \leq \sqrt{P_j^{max}}& \mbox{\;\;\;} j = 1,\ldots, N_r \IEEEnonumber \\
  & \bOmega - \bw \bw^H \succeq 0. \IEEEnonumber
\end{IEEEeqnarray}
 The feasibility check problem can be solved by employing Second Order Cone Programming (SOCP) using cvx software \cite{cvx}. It is worth emphasizing that no rank deduction algorithm is required to obtain a rank one solution for $\bOmega$. In the next section, we define and solve the QoS's maximization problem when any imperfect interference CSI is available to the central controller.

\section{SINR's Optimization with Imperfect Interferers CSI}\label{Sec:ImperfectCSI}
In session \ref{Sec:PerfectCSI}, the problem of joint optimizing the SINR of the PUs and SUs considering the perfect knowledge of the external interferences is discussed. However, the assumption of perfect knowledge on the unfriendly interferers' CSI is not realistic in practical applications. We provide an optimization problem which is robust against all uncertainties in interferers' CSI.
Let $\Psi$ denote the set of uncertainty regions of all interferers' CSI. By using (\ref{Equ:ChannelIntBounded}) we have:
\begin{IEEEeqnarray}{rCl}\label{Def:Psi}
  \Psi &=& \left\{ \forall \left\{\{\nabla\bh_{P_i}\}_{i = 1,2},\{\nabla\bh_{S_i}\}_{i = 1,2},\{\nabla\bh_{I_l}\}_{l = 1,\ldots,N_I} \right\}
  \left| \| \nabla\bh_{P_i} \| \leq \epsilon_{P_i}  \&  \| \nabla\bh_{S_i} \| \leq \epsilon_{S_i} \&  \| \nabla\bh_{I_l} \| \leq \epsilon_{l} \right.\right\}\mbox{\;} \IEEEeqnarraynumspace \IEEEyesnumber
\end{IEEEeqnarray}
 By applying (\ref{Def:Psi}), the robust version of the optimization problem (\ref{Equ:Optimizationwgamma}) can be written as:
 \begin{IEEEeqnarray}{rCl}\label{Equ:OptimizationwgammaImperfect}
  \max_{\bw, \gamma>0} & \gamma &\IEEEyesnumber
\IEEEnosubnumber\\
  \mbox{Subject To:}& SINR_{P_i}(\psi) \geq \gamma, & \mbox{\;\;\;}\forall \psi \in \Psi \mbox{\;\;\;} i=1,2 \IEEEnonumber \\
  &\mu SINR_{S_i}(\psi) \geq \gamma, & \mbox{\;\;\;} \forall \psi \in \Psi  \mbox{\;\;\;} i=1,2 \IEEEnonumber \\
   & P_{r_j}(\psi) \leq P_j^{max}, & \mbox{\;\;\;} \forall \psi \in \Psi \mbox{\;\;\;} j = 1,\ldots,N_r  \IEEEnonumber.
\end{IEEEeqnarray}
 The robust optimization problem (\ref{Equ:OptimizationwgammaImperfect}) suggests common $\psi$ for all constraints. However, one may think of a separate uncertainty array $\psi$ for every set of constraints. In fact, Theorem 2.1 of \cite{ben1998robust} allows us to write a counterpart robust optimization problem of (\ref{Equ:OptimizationwgammaImperfect}) as:
\begin{IEEEeqnarray}{rCl}\label{Equ:OptimizationwgammaImperfectCounterpart}
  \max_{\bw, \gamma>0} & \gamma &\IEEEyesnumber
\IEEEnosubnumber\\
  \mbox{Subject To:}& SINR_{P_i}(\psi_{P_i}) \geq \gamma, & \mbox{\;\;\;}\forall \psi_{P_i} \in \Psi  \mbox{\;\;\;} i=1,2 \IEEEnonumber \\
  &\mu SINR_{S_i}(\psi_{S_i}) \geq \gamma, & \mbox{\;\;\;}\forall \psi_{S_i} \in \Psi \mbox{\;\;\;} i=1,2 \IEEEnonumber \\
   & P_{r_j}(\psi_{r_j}) \leq P_j^{max}, & \mbox{\;\;\;}\forall \psi_{r_j} \in \Psi \mbox{\;\;\;} j = 1,\ldots,N_r \IEEEnonumber,
\end{IEEEeqnarray}
or equivalently:
\begin{IEEEeqnarray}{rCl}\label{Equ:OptimizationwgammaImperfectworst}
  \max_{\bw, \gamma>0} & \gamma  \IEEEyesnumber
\IEEEnosubnumber\\
  \mbox{Subject To:}&\min_{\forall \psi_{P_i} \in \Psi} SINR_{P_i}(\psi_{P_i}) \geq \gamma, & \mbox{\;\;\;} i = 1,2 \IEEEeqnarraynumspace \IEEEyessubnumber\\
  &\mu \min_{ \forall \psi_{S_i} \in \Psi}SINR_{S_i}(\psi_{S_i}) \geq \gamma, & \mbox{\;\;\;} i=1,2 \IEEEeqnarraynumspace \IEEEyessubnumber \\
   &\max_{\forall \psi_{r_j} \in \Psi} P_{r_j}(\psi_{r_j}) \leq P_j^{max},  & \;\;\; j = 1,\ldots,N_r  \IEEEyessubnumber .
\end{IEEEeqnarray}
In the next subsection, a closed-form format for all constraints of the optimization problem (\ref{Equ:OptimizationwgammaImperfectworst}) is provided.
\subsection{Closed-Form Worst Case Imperfection}\label{Subsec:WorstClosedForm}
By using the definitions (\ref{Equ:ChannelIntBounded}), (\ref{Def:Q}) and (\ref{Def:zeta}), the constraints (\ref{Equ:OptimizationwgammaImperfectworst}a-b) can be rewritten as:
\begin{IEEEeqnarray}{rCl}\label{Equ:SINRConstworst}
& \gamma \max_{\forall \psi_{P_i} \in \Psi} \left\{\sum_{l=1}^{N_I}{P_{I_l}}\| \bw^H\bF_{P_i}(\hat{\bh_{I_l}}+ \nabla\bh_{I_l})\|^2
+  \|(\hat{\bh_{P_i}} + \nabla\bh_{P_i})^H\bP_I^{\frac{1}{2}}\|^2 \right\} \IEEEnonumber\\
&\leq\bw^H (P_{P_{\bar{i}}} \bk_{P_1P_2}\bk_{P_1P_2}^H -  \gamma \bT_{P_i}) \bw  - \gamma \sigma^2  \IEEEyesnumber \IEEEyessubnumber \\
  & {\gamma} \max_{\forall \psi_{S_i} \in \Psi} \left\{\sum_{l=1}^{N_I}{P_{I_l}}\| \bw^H\bF_{S_i}(\hat{\bh_{I_l}}+ \nabla\bh_{I_l})\|^2 +  \|(\hat{\bh_{S_i}} + \nabla\bh_{S_i})^H\bP_I^{\frac{1}{2}}\|^2 \right\} \IEEEnonumber\\
  &\leq\bw^H (\mu P_{S_{\bar{i}}}\bk_{S_1S_2}\bk_{S_1S_2}^H - \gamma \bT_{S_i})\bw - \gamma \sigma^2 \IEEEyessubnumber
\end{IEEEeqnarray}
where $\bP_I = [P_{I_1},P_{I_1},\ldots,P_{I_{N_I}}]^T$,
and
\begin{IEEEeqnarray}{rCl}\label{Def:R}
 \bT_{P_i} &=&  \sum_{j=1}^{2}{P_{S_j}\bk_{S_jP_i}\bk_{S_jP_i}^H}+\sigma^2 \bF_{P_i} \bF_{P_i}^H, \;\;\; i = 1,2  \IEEEyesnumber \IEEEyessubnumber\\
 \bT_{S_i} &=& \sum_{j=1}^{2}{P_{P_j}\bk_{S_iP_j}\bk_{S_iP_j}^H}+\sigma^2 \bF_{S_i} \bF_{S_i}^H, \;\;\; i = 1,2.
  \IEEEyessubnumber
\end{IEEEeqnarray}

Also, by using (\ref{Equ:ChannelIntBounded}) and (\ref{Equ:Pi_individual_simple}) the constraint (\ref{Equ:OptimizationwgammaImperfectworst}c) can be written as:
\begin{IEEEeqnarray}{rCl}\label{Equ:PowerConstWorst}
  &\max_{\forall \psi_{r_j} \in \Psi}\{ \sum_{l=1}^{N_I}{{P_{I_l}}|{\boldsymbol{\ell}_{N_r,j}}^T(\hat{\bh}_{I_l}+\nabla {\bh}_{I_l})|^2}\} | {\boldsymbol{\ell}_{N_r,j}}^T \bw|^2 \leq P_j^{max} - \chi_{r_j} | {\boldsymbol{\ell}_{N_r,j}}^T \bw|^2\IEEEyesnumber
\end{IEEEeqnarray}
where,
\begin{IEEEeqnarray}{rCl}\label{Def:chi}
\chi_{r_j} =\sum_{i=1}^{2}{{P_{P_i}}|{\boldsymbol{\ell}_{N_r,j}}^T\bof_{P_i}|^2}
     +\sum_{i=1}^{2}{{P_{S_i}}|{\boldsymbol{\ell}_{N_r,j}}^T \bof_{S_i}|^2}+\sigma^2,\;\;\; j = 1,\ldots,N_r
\end{IEEEeqnarray}
In order to write the constraints (\ref{Equ:SINRConstworst}) and (\ref{Equ:PowerConstWorst}) in a closed format, we use two lemmas on norm vector inequalities. First:
\begin{lemma}\label{Lemma:One}
  Let $\bc$ and $\ba$ denote complex vectors. If $\bb$ is a vector whose norm is bounded by constant $\epsilon$, i.~e.~ $\|\bb\| \leq \epsilon$, then $\|\bc^H(\ba + \bb)\| \leq |\bc^H\ba| + \epsilon \|\bc\|$ and the equality holds if and only if  $\bb = \frac{\epsilon}{\|\bc\|}\bc e^{j\angle{\bc^H\ba}}$.
\end{lemma}
\begin{proof}
  The proof is provided in \cite{CRBidirectional-zaeri2016} by using triangle inequality theorem and Cauchy-Schwarz inequality.
\end{proof}
By applying lemma \ref{Lemma:One} we have:
\begin{IEEEeqnarray}{rCl}\label{Equ:UncertainwPSR}
  \max_{\forall \psi_{P_i} \in \Psi} &\| \bw^H\bF_{P_i}(\hat{\bh}_{I_l}+ \nabla\bh_{I_l})\|  = |\bw^H\bF_{P_i}\hat{\bh}_{I_l}|+ \epsilon_l \|\bw^H\bF_{P_i}\| \IEEEyesnumber
   \IEEEyessubnumber \\
   \max_{\forall \psi_{S_i} \in \Psi} &\| \bw^H\bF_{S_i}(\hat{\bh}_{I_l}+ \nabla\bh_{I_l})\|  =  |\bw^H\bF_{S_i}\hat{\bh}_{I_l}|+ \epsilon_l \|\bw^H\bF_{S_j}\| \IEEEyesnumber
   \IEEEyessubnumber \\
    & \max_{\forall \psi^r_j \in \Psi} \mbox{\;\;\;} |{\boldsymbol{\ell}_{N_r,j}}^T(\hat{\bh}_{I_j}+\nabla {\bh}_{I_j})|
   =  |{\boldsymbol{\ell}_{N_r,j}}^T \hat{\bh}_{I_j}|+ \epsilon_{l}  \IEEEyesnumber \IEEEyessubnumber
\end{IEEEeqnarray}
Next,
\begin{lemma}\label{Lemma:Two}
  Let $\bDelta$ denote an invertible matrix, let $\ba$ represent a vector and let $\bb$ represent a vector whose norm is upper bounded by constant $\epsilon$, i.~e.~ $\|\bb\| \leq \epsilon$. Then $\|(\ba+\bb)^H \bDelta \| \leq (1 + \frac{\epsilon}{\|\ba\|})\|\ba^H \bDelta\|$ and the equality holds if and only if $\bb = \frac{\epsilon}{\|\ba\|}\ba$.
\end{lemma}
\begin{proof}
  The proof is provided in \cite{CRBidirectional-zaeri2016} by using triangle inequality theorem and Cauchy-Schwarz inequality.
\end{proof}
By defining:
\begin{IEEEeqnarray}{rCl}\label{Def:kappa}
\kappa_{p_i} = (1+\frac{\epsilon_{P_i}}{\| \hat{\bh_{P_i}}\|})^2\| \hat{\bh_{P_i}}^H\bP_I^{\frac{1}{2}}\|^2  \qquad \mbox{and} \qquad
\kappa_{s_i} = (1+\frac{\epsilon_{S_i}}{\|\hat{\bh_{S_i}}\|})^2\| \hat{\bh_{S_i}}^H\bP_I^{\frac{1}{2}}\|^2, \;\;\; i = 1,2,
\end{IEEEeqnarray}
and by using Lemma \ref{Lemma:Two}, we have:
\begin{IEEEeqnarray}{rCl}\label{Equ:UncertainPS}
   \max_{\forall \psi_{P_i} \in \Psi} &\|(\hat{\bh_{P_i}} + \nabla\bh_{P_i})^H\bP_I^{\frac{1}{2}}\|^2 = \kappa_{p_i} \qquad \mbox{and} \qquad
   \max_{\forall \psi_{S_i} \in \Psi} & \|(\hat{\bh_{S_i}} + \nabla\bh_{S_i})^H\bP_I^{\frac{1}{2}}\|^2  = \kappa_{s_i}, \;\;\; i = 1,2.\IEEEyessubnumber
\end{IEEEeqnarray}
As a result, the constraints (\ref{Equ:SINRConstworst}) and (\ref{Equ:PowerConstWorst}) can be rewritten as the following closed format:
\begin{IEEEeqnarray}{rCl}\label{Equ:closedConst}
  & \gamma \left(\sum_{l=1}^{N_I}{P_{I_l}} (|\bw^H\bF_{P_i}\hat{\bh}_{I_l}|+ \epsilon_l \|\bw^H\bF_{P_i}\|)^2 +\kappa_{p_i} \right) \leq
   \bw^H (P_{P_{\bar{i}}} \bk_{P_1P_2}\bk_{P_1P_2}^H -  \gamma \bT_{P_i}) \bw  - \gamma \sigma^2, \;\;\; i = 1,2  \IEEEyesnumber \IEEEyessubnumber \\
  & {\gamma} \left(\sum_{l=1}^{N_I}{P_{I_l}}(|\bw^H\bF_{S_i}\hat{\bh}_{I_l}|+ \epsilon_l \|\bw^H\bF_{S_j}\|)^2+ \kappa_{s_i}\right) \leq
   \bw^H (\mu P_{S_{\bar{i}}}\bk_{S_1S_2}\bk_{S_1S_2}^H - \gamma \bT_{S_i})\bw - \gamma \sigma^2, \;\;\; i = 1,2 \IEEEyessubnumber \\
  & \kappa_{r_j} | {\boldsymbol{\ell}_{N_r,j}}^T \bw|^2  \leq P_j^{max}, \;\;\; j = 1,\ldots, N_r\IEEEyessubnumber
\end{IEEEeqnarray}
where
\begin{IEEEeqnarray}{rCl}\label{Def:kappar}
\kappa_{r_j} =  \sum_{l=1}^{N_I}{{P_{I_l}}(|{\boldsymbol{\ell}_{N_r,j}}^T \hat{\bh}_{I_l}|+ \epsilon_{l})^2}  + \chi_{r_j},\;\;\; j = 1,\ldots, N_r
\end{IEEEeqnarray}
The closed-form constraints are substituted in the optimization problem (\ref{Equ:OptimizationwgammaImperfectworst}). In the next subsection, we use SOCP method to obtain the optimal beamforming of our system model considering the imperfection of interferers' CSI.
\subsection{Optimal SINR Solution}\label{Subsec:BisectionImPerfect}
In this subsection, we provide a bisection method to find the optimal value of $\gamma$ when just imperfect interferers' CSI are known. By applying (\ref{Equ:closedConst}), the optimization problem (\ref{Equ:OptimizationwgammaImperfectworst}) is turned to a feasibility check problem of finding $\bw$ with the constraints (\ref{Equ:closedConst}) for a given value of $\gamma$. The upper bound on $\gamma$ for the case when perfect knowledge of interferers' CSI is available was described in (\ref{Equ:Uppergamma}). It should be noted that the same upper bound applies for the case when only an imperfect knowledge of interferers' CSI is available. Therefore, the same bisection approach as is used in subsection \ref{Subsec:BisectionPerfect} can be applied for the imperfection case. Drawing to a close, we modify the constraints (\ref{Equ:closedConst}) to the SOCP format, so the discussed feasibility check problem is solved in an efficient way.

Let us use auxiliary positive relay variables $\rho_{p_i}$ and $\rho_{s_i}$, $i = 1,2$ in such a way that $\|\bw^H\bF_{P_i}\| \leq \rho_{p_i}$ and $\|\bw^H\bF_{S_i}\| \leq \rho_{s_i}$.
Also, we use the $N_I \times 1$ auxiliary positive real vector variables $\boldsymbol{\varrho}_{p_i} = [\varrho_{p_i,1}, \ldots , \varrho_{p_i,N_I}]^T$ and $\boldsymbol{\varrho}_{s_i} = [\varrho_{s_i,1}, \ldots , \varrho_{s_i,N_I}]^T$ for $i = 1,2$ in such a way that $|\bw^H\bF_{P_i}\hat{\bh}_{I_l}| \leq \varrho_{p_i,l}$ and $|\bw^H\bF_{S_i}\hat{\bh}_{I_l}| \leq \varrho_{s_i,l}$ for $l = 1,\ldots,N_I$.
By using the above mentioned auxiliary variables the feasibility check problem for a given value of $\gamma$ can be written as:
\begin{IEEEeqnarray}{rCl}\label{Equ:FindwImperfect}
& \mbox{Find} \;\;\; { \{\rho_{p_i},\rho_{s_i},\boldsymbol{\varrho}_{p_i},\boldsymbol{\varrho}_{s_i}\}_{i = 1,2} ,\bw} \IEEEyesnumber \\
 & \gamma \left(\sum_{l=1}^{N_I}{P_{I_l}} ({\boldsymbol{\ell}_{N_I,l}}^T \boldsymbol{\varrho}_{p_i} + \epsilon_l \rho_{p_i})^2 +\kappa_{p_i}\right)\leq
   \bw^H (P_{P_{\bar{i}}} \bk_{P_1P_2}\bk_{P_1P_2}^H -  \gamma \bT_{P_i}) \bw  - \gamma \sigma^2,  & \mbox{\;\;\;} i = 1,2\IEEEyessubnumber \\
  & {\gamma} \left(\sum_{l=1}^{N_I}{P_{I_l}}({\boldsymbol{\ell}_{N_I,l}}^T \boldsymbol{\varrho}_{p_i} + \epsilon_l \rho_{p_i})^2 + \kappa_{s_i}  \right) \leq
   \bw^H (\mu P_{S_{\bar{i}}}\bk_{S_1S_2}\bk_{S_1S_2}^H - \gamma \bT_{S_i})\bw - \gamma \sigma^2, & \mbox{\;\;\;} i = 1,2\IEEEyessubnumber \\
  &\sqrt{\kappa_{r_j}} | {\boldsymbol{\ell}_{N_r,j}}^T \bw|  \leq \sqrt{P_j^{max}}, & \mbox{\;\;\;} j = 1,\ldots,N_r  \IEEEyessubnumber \\
  &\|\bw^H\bF_{P_i}\| \leq \rho_{p_i} \mbox{\;\;} \qquad , \qquad \|\bw^H\bF_{S_i}\| \leq \rho_{s_i} &\mbox{\;\;\;} i = 1,2\IEEEyessubnumber \\
  &|\bw^H\bF_{P_i}\hat{\bh}_{I_l}| \leq {\boldsymbol{\ell}_{N_I,l}}^T \boldsymbol{\varrho}_{p_i} \qquad , \qquad
  |\bw^H\bF_{S_i}\hat{\bh}_{I_l}| \leq {\boldsymbol{\ell}_{N_I,l}}^T \boldsymbol{\varrho}_{s_i}, &\mbox{\;} i = 1,2 ~\mbox{and} ~   l = 1,\ldots,N_I ~ \IEEEeqnarraynumspace \IEEEyessubnumber.
\end{IEEEeqnarray}
All (\ref{Equ:FindwImperfect}c-e) constraints represent SOC regions.
We use auxiliary vector variables $\boldsymbol{\varpi}_{p_i} = [{\varpi}_{p_i,1},\ldots,{\varpi}_{p_i,N_I}]^T$ and $\boldsymbol{\varpi}_{s_i} = [{\varpi}_{s_i,1},\ldots,{\varpi}_{s_i,N_I}]^T$, in the interest of clarity, in which ${\varpi}_{p_i,l} = {\boldsymbol{\ell}_{N_I,l}}^T \boldsymbol{\varrho}_{p_i} + \epsilon_l \rho_{p_i}$ and ${\varpi}_{s_i,l} =  {\boldsymbol{\ell}_{N_I,l}}^T \boldsymbol{\varrho}_{s_i} + \epsilon_l \rho_{s_i}$ for $i = 1,2$ and $ l = 1,\ldots,N_I$.

By assuming that $\bk_{P_1P_2}^H \bw$ is a positive real number, the constraints (\ref{Equ:FindwImperfect}a) can be turned to the following SOC constraint:
\begin{IEEEeqnarray}{rCl}\label{Equ:SOCPconstImperfect}
 &  \frac{P_{P_{\bar{i}}}}{\gamma} \bk_{P_1P_2}^H \bw \geq  \sqrt{\boldsymbol{\varpi}_{p_i}^T \bP_I \boldsymbol{\varpi}_{p_i} + \bw^H \bT_{P_i} \bw +\kappa_{p_i} +\sigma^2},\;\;\; i = 1,2.  \IEEEyesnumber.
\end{IEEEeqnarray}

Ultimately, by using Lemma \ref{Lemma:QOP2SOCP} and by defining matrices $\bP_{s_i} = \textsl{blkdiag}(\bP_I,\bT_{S_i})$, $i = 1,2$, the constraints (\ref{Equ:FindwImperfect}b) can be relaxed to the following SOC format:
\begin{IEEEeqnarray}{rCl}
  & \left\| \left[\begin{array}{c}
                    1 - \frac{\mu P_{S_{\bar{i}}}}{\gamma}   \bk_{S_1S_2}\bk_{S_1S_2}^H \bullet \bOmega+ \kappa_{s_i} +\sigma^2 \\
                    2 \bP_{s_i}^{\frac{1}{2}}\left[\begin{array}{c}
                                                     \boldsymbol{\varpi}_{s_i} \\
                                                     \bw
                                                   \end{array} \right]
                  \end{array}\right]\right\|
\leq 1 + \frac{\mu P_{S_{\bar{i}}}}{\gamma}  \bk_{S_1S_2}\bk_{S_1S_2}^H \bullet \bOmega - \kappa_{s_i} - \sigma^2, \;\;\; i = 1,2.\IEEEyesnumber
\end{IEEEeqnarray}
 Therefore, the feasibility check problem (\ref{Equ:FindwImperfect}) can be effectively solved using the SOCP method.

\section{Numerical Experiments}\label{Sec:Simulation}
\begin{figure}
\centerline{\resizebox{!}{8.5cm}{\includegraphics{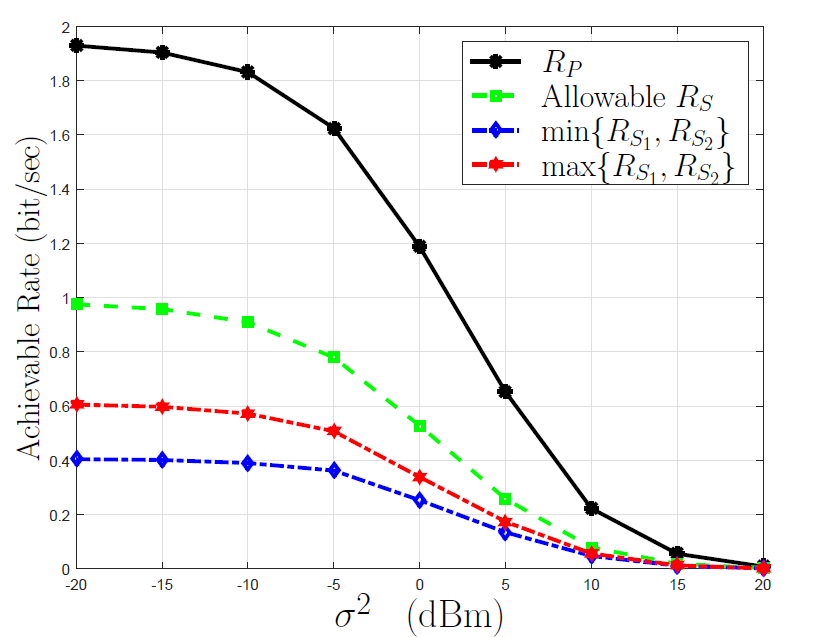}}}
\caption{Achievable rate for 1 Hz bandwidth for primary and secondary transceivers for fairness design parameter $\mu = 3$.}    \label{fig:SINRPUSUmu3Compare}
 \end{figure}

In this section, we present different simulation scenarios to illustrate the effectiveness of our proposed method in combating interference when the perfect information on CSIs of channels between the interferers and other users is not available. To do so, all channel coefficients were generated as complex Gaussian variables with zero mean and unit variance. The PU and SU transceivers' powers were assumed to be 0 dBm. For the sake of simplicity, we considered two unfriendly interferers that operate in the PUs' spectrum property, with the power of -1 dBm. Also, 10 relays were considered to cooperate with both PU and SU transceivers with a maximum individual power limit of 1 dBm. We consider 3 times priority for the primaries quality of service with respect to the secondaries, i.e., $\mu = 3$. The achievable rate at each receiver is defined by $R = \log_2(1 + SINR)$ and is plotted in Fig.~\ref{fig:SINRPUSUmu3Compare} versus the noise power ($\sigma^2$). The minimum achievable rate of $PU_1$ and $PU_2$ is shown by $R_P$ and plotted versus $\sigma^2$. As can be seen in the figure, $R_P$ decreased from 2 to 0 when noise power increased from -20 to 20 dBm. Also, by using the priority design parameter $\mu = 3$, the optimization problem forces the network to assign more resources for primary transmissions purposes. This limits the achievable rate for the secondary transmission as it is shown by allowable $R_S$ in the figure. As it can be seen in Fig. \ref{fig:SINRPUSUmu3Compare}, the optimization problem makes a restriction on maximum achievable rate of 1 when $\sigma^2$ is -20 dBm for $R_{S_1}$ and $R_{S_2}$. However, at this level of noise power, $R_{S_1}$ and $R_{S_2}$ achieve the rates of 0.4 and 0.6 dBm, respectively.


\begin{figure}
\centerline{\resizebox{!}{8.5cm}{\includegraphics{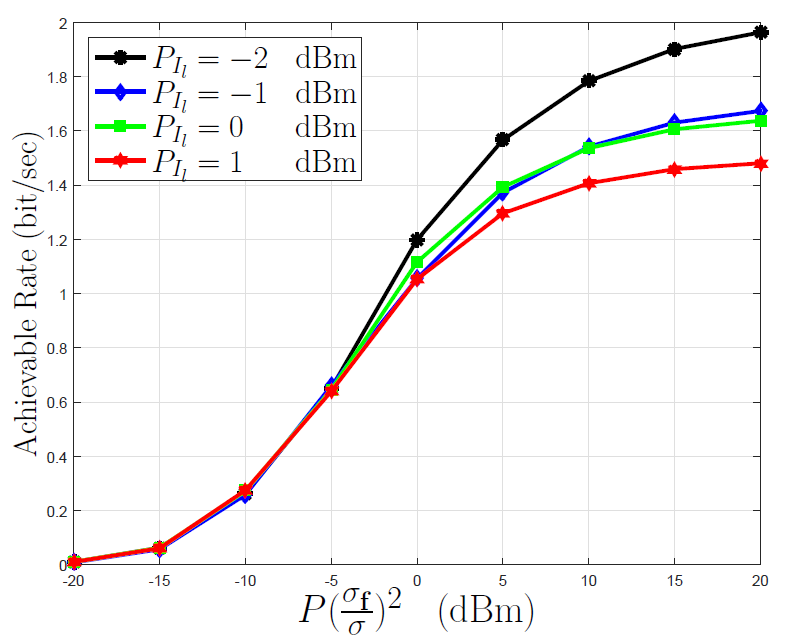}}}
    \caption{Achievable rate for 1 Hz bandwidth versus $P (\frac{\sigma_{\bof}}{\sigma})^2$ for different interferers power.}
     \label{fig:SINRIntPower}
 \end{figure}

The impact of the interferers on the SDR system is illustrated in Fig.~ \ref{fig:SINRIntPower} for different transmission powers of these interferers. The designed parameter $\mu$ was considered to be 1, and achievable rate were obtained for different levels of interferers' power, while all other parameters were the same as before. Figure \ref{fig:SINRIntPower} shows that, by increasing the interferers' power from -2 dBm to 1 dBm, the rate decreased by 2.5 dB when the noise is too weak. However, this performance reduction is regulatable if noise variance is high.

%

\begin{figure}
\centerline{\resizebox{!}{8.5cm}{\includegraphics{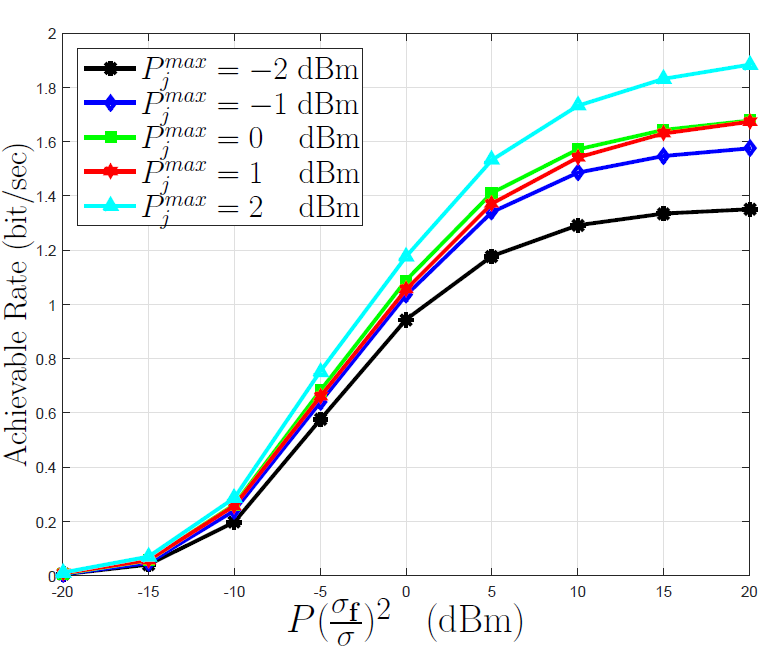}}}
 \caption{Achievable rate for 1 Hz bandwidth versus $P (\frac{\sigma_{\bof}}{\sigma})^2$ for different maximum limitation on individual relay powers.}
   \label{fig:SINRRelayPower}
 \end{figure}

We also investigated the effects of cooperation level of relays on interference mitigation. In this scenario, we considered 2 interferers and, the maximum power that each relay was assigned for relaying purposes, was varied from -2 to 2 dBm while all other parameters were the same as before. Figure \ref{fig:SINRRelayPower} shows the changes on the average $\gamma$ in a Monte-Carlo simulation versus the noise power for different maximum limitations on individual relay powers. As it is shown in this figure, by increasing the relay power limit, the diversity gain of the system increased. For instance, increasing relay power limit from -2 dBm to 2 dBm caused a 3 dB increase on SINR when the noise power is -20 dBm.

\begin{figure}
\centerline{\resizebox{!}{8.5cm}{\includegraphics{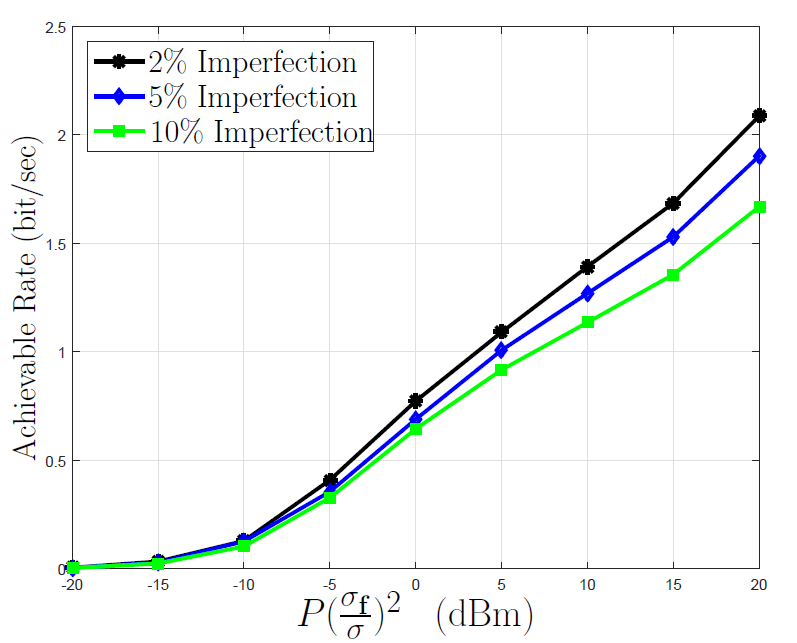}}}
\caption{Achievable rate for 1 Hz bandwidth versus $P (\frac{\sigma_{\bof}}{\sigma})^2$ for different level of imperfection in knowledge of interferers' CSI.}
   \label{fig:ImperfectPercentage}
 \end{figure}

The previous numerical experiments were based on the assumption of perfect CSI availability for interferes. We examined our proposed robust method against uncertainties on interferers' CSI. A scenario where the interferers' CSI are known imperfectly is considered. We assumed the imperfection as a percentage of the estimated CSI and varied this percentage from $2\%$ to $10\%$. The optimal robust $\gamma$ is calculated by solving the robust optimization problem (\ref{Equ:FindwImperfect}). Figure \ref{fig:ImperfectPercentage} illustrates that the system will pay the cost by gaining less SINR if the knowledge accuracy is decreased. For example, the performance of the system decreased by 1 dB if the accuracy of the interferers' CSI decreases from 2 to 10 percent.

\begin{figure}
\centerline{\resizebox{!}{8.5cm}{\includegraphics{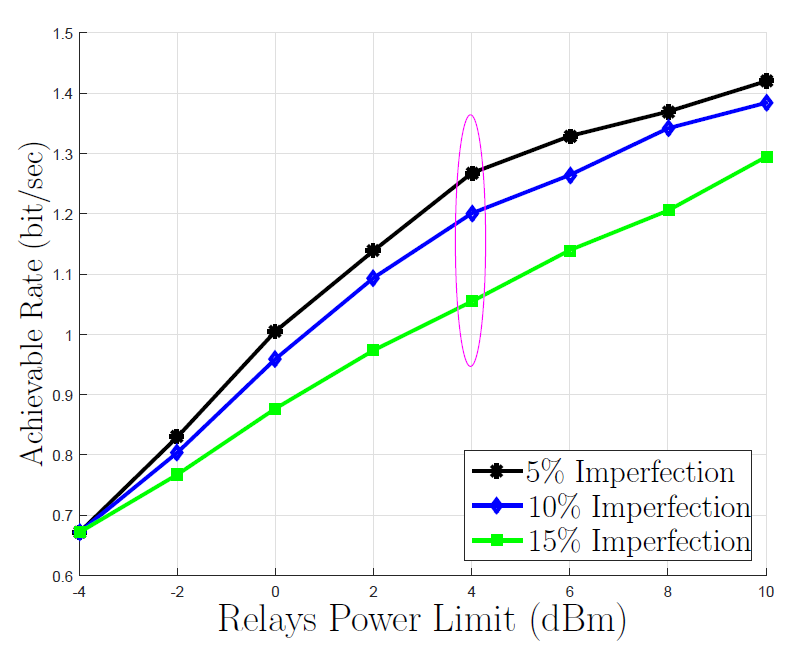}}}
\caption{Achievable rate for 1 Hz Bandwidth versus relays' power limit for different level of imperfection in knowledge of interferers' CSI.}
  \label{fig:ImperfectPercentageRelayP}
 \end{figure}

The $\sigma^2$ value is fixed on 0 dB in Fig.~\ref{fig:ImperfectPercentageRelayP} and the achievable rate is plotted versus the changes in the individual relay powers for different levels on the interferers CSI imperfection. As it is mentioned before, the relays provide diversity gain to the system and the achievable rates for the transceivers in the system increase when the relays power increase. Besides, more uncertainty on the interferers CSI causes less performance for the system. As it can be seen from this figure, the achievable rate decreases from 1.25 to 1.05 when the imperfection on CSI increased between 5 and 15 percentages for a maximum available power of 4 dBm for the relays.

To summarize, in the simulations, we showed the diversity gain on a two-way SDR-network in which the resources are fairly distributed among primary and secondary users. The effect of interferers in our cooperative model was investigated. Moreover, the interferers CSI uncertainties effect on the system performance was investigated and it was shown that our proposed method is robust against such uncertainties. It is worth mentioning that the convergence of the proposed method depends on the convergence of the bisection method which is in order of $O(\frac{1}{2^n})$ \cite{numerical_book}.  While the bisection method may be slower than numerical methods, but it always converges to the solution.

\section{Conclusion}\label{Sec:Conclusion}
 We propose a model for co-existence different types of secondary users in an SDR network. A system consists of two pairs of PU and SU transceivers, several SU relays and also several interferers is considered. The optimum beamforming solution is provided to maximize quality service in PU and SU transceivers. Moreover, the optimization problem is solved by considering the worst-case scenario when the knowledge on the interferers channels is imperfect. The simulation results show the performance of our proposed method and its robustness against uncertainties in interferers CSI.

\bibliographystyle{IEEEtran}
\bibliography{IEEEabrv,reference}
\appendix \label{Appendix:One}

\end{document}